\documentclass[showpacs,twocolumn,preprintnumbers,amsmath,amssymb]{revtex4}
\usepackage{amsmath,amsfonts,latexsym,amssymb,graphicx,graphics,epsfig,subfigure,color,makeidx}
\usepackage{xcolor,diagbox}
\usepackage{multirow}
\usepackage[colorlinks,linkcolor=blue,anchorcolor=blue,citecolor=green,urlcolor=blue]{hyperref}
\newcommand {\nn}{\nonumber}

\begin{document}
\title{Gravitational resonances on $f(T)$-branes}

\author{Qin Tan$^{a}$$^{c}$}
\author{Wen-Di Guo$^{a}$$^{c}$$^{d}$}
\author{Yu-Peng Zhang$^{a}$$^{c}$}
\author{Yu-Xiao Liu$^{a}$$^{b}$$^{c}$\footnote{liuyx@lzu.edu.cn, corresponding author}}

\affiliation{$^{a}$Institute of Theoretical Physics $\&$ Research Center of Gravitation, Lanzhou University, Lanzhou 730000, China\\
	$^{b}$Key Laboratory for Magnetism and Magnetic of the Ministry of Education, Lanzhou University, Lanzhou 730000, China\\
	$^{c}$Joint Research Center for Physics, Lanzhou University and Qinghai Normal University, Lanzhou 730000, China\\
	$^{d}$CENTRA, Departamento de F\'{\i}sica, Instituto Superior T\'ecnico -- IST, Universidade de Lisboa -- UL, Avenida Rovisco Pais 1, 1049 Lisboa, Portugal\\}

\begin{abstract}
	In this work, we investigate the gravitational resonances in various $f(T)$-brane models with the warp factor $\text{e}^{A(y)}=\tanh\big(k(y+b)\big)-\tanh\big(k(y-b)\big)$, where $f(T)$ is an arbitrary function of the torsion scalar $T$. For three kinds of $f(T)$, we give the solutions to the system. Besides, we consider the tensor perturbation of vielbein and obtain the effective potentials by the Kaluza-Klein (KK) decomposition. Then, we analyze what kind of effective potential can produce the gravitational resonances. Effects of different parameters on the gravitational resonances are analysed. The lifetimes of the resonances could be long enough as the age of our universe in some ranges of the parameters. This indicates that the gravitational resonances might be considered as one of the candidates for dark matter. Combining the current experimental observations, we constrain the parameters for these brane models.
\end{abstract}

\maketitle

\section{Introduction}
\label{Introduction}

It is known that extra-dimensional theories have been developed for about one hundred years. In 1920s, in order to unify electromagnetic interaction and gravitational interaction, Kaluza and Klein (KK) proposed a five-dimensional spacetime theory by introducing a compact spatial dimension with Planck scale~\cite{kaluza1921unitatsproblem,Klein:1926tv}. In 1982, Keiichi Akama presented the ¡°braneworld¡± model which picture that we live in a dynamically localized 3-brane in higher-dimensional spacetime~\cite{Akama:1982jy}. In 1983, Rubakov and Shaposhnikov proposed the domain wall model in a five-dimensional flat spacetime~\cite{Rubakov:1983bb}. In this model, the domain wall is generated by a scalar field with a kink configuration and the extra dimension can be infinite. Fermions can be localized on the domain wall by a Yukawa coupling. However, the four-dimensional effective Newtonian gravity cannot be recovered from this model. Then, more than twenty years ago, in order to solve the huge hierarchical problem between the weak and Planck scales, some brane world models were presented. The two famous ones are the large extra dimension model proposed by Arkani-Hamed, Dimopoulos, and Dvali (ADD)~\cite{ArkaniHamed:1998rs} and the warped extra dimension model (RS-I) by Randall and Sundrum (RS) \cite{Randall:1999ee}. In these brane models, the sizes of extra dimensions are finite. A great development was achieved in Ref.~\cite{Randall:1999vf}, which shows that the four-dimensional gravity can be recovered on the brane even though the extra dimension is  infinite. After that, extra-dimensional theories attracted a lot of interests~\cite{Goldberger:1999uk,Gremm:1999pj,DeWolfe:1999cp,Bazeia:2008zx,Charmousis:2001hg,Arias:2002ew,Barcelo:2003wq,
	Bazeia:2004dh,CastilloFelisola:2004eg,Kanno:2004nr,BarbosaCendejas:2005kn,Koerber:2008rx,BarbosaCendejas:2007vp,
	Johnson:2008kc,Almeida:2009jc,Liu:2011wi,Chumbes:2011zt,Andrianov:2012ae,Kulaxizi:2014yxa,Dutra:2014xla,Karam:2018squ}.

In this paper, we are interested in thick brane models. In most of these models, branes are generated dynamically by one or more background scalar fields~\cite{Dzhunushaliev:2009va,Navarro:2004di,deSouzaDutra:2008gm,Zhong:2014kha,Xie:2019jkq,Zhou:2017xaq}. Various matter fields and gravity in the higher-dimensional spacetime should have the ability to explain the physics in the four-dimensional spacetime. Therefore, in order to recover the standard model and the effective four-dimensional Newtonian potential, the zero modes of these matter fields and tensor fluctuations of gravity should be localized on branes~\cite{Csaki:2000fc}.
In addition to the zero modes, we will get massive KK modes, which are new particles predicted by these theories. Generally, for the case of a thick brane embedded in five-dimensional asymptotic Anti-de Sitter (AdS) spacetime, the effective potential felt by KK modes along the extra dimension is volcano-like. In this case, the massive KK modes cannot be localized on the brane, but a finite number of massive KK modes could be quasi-localized on the brane~\cite{{Liu:2009ve}}. These quasi-localized KK modes are called resonant KK modes. We can judge whether there are resonances by analyzing the shape of the supersymmetric partner potential of the effective potential \cite{Cooper:1994eh}. In this paper, we focus on the case of gravitational resonant KK modes, which also contribute to the four-dimensional Newtonian potential~\cite{Cruz:2013uwa,Xu:2014jda,Csaki:2000pp}. This provides a possible way to detect the extra dimension. In fact, in the Gregory-Rubakov-Sibiryakov (GRS) model, the four-dimensional Newtonian gravitational force is generated by the quasilocalized gravitons~\cite{Gregory:2000jc}. Furthermore, if the lifetime of the resonance can be long enough as the age of our universe, they might be a candidate for dark matter~\cite{Sui:2020fty}. Therefore, in braneworld models, the investigation of gravitational resonances is an important topic. In other scenarios, there are also gravitational resonances, such as the quasinormal modes of black hole fluctuation theory~\cite{Nollert:1999ji}.

It is well known that general relativity (GR) is a theory of gravity with only curvature. In 1928, Einstein established a gravitational theory with only torsion in spacetime in order to unify gravitational interaction and electromagnetic interaction, called Teleparallel equivalence of general relativity (TEGR)~\cite{Hayashi:1979qx,Sousa:2007zc}. It is in fact equivalent to GR, based on the fact that torsion scalar $T$ differs from the Ricci scalar $R$ only by a boundary term. Although the field equation of TEGR is equivalent to GR, the spacetime geometry described by TEGR is different from GR. In TEGR, the dynamic field is vielbein, which is defined in the tangent space at each point in spacetime. Inspired by $f(R)$ gravity theory, Bengochea and Ferraro firstly proposed the generalization of teleparallel gravity, $f(T)$ gravity ($T$ is the torsion scalar and $f(T)$ is an arbitrary function of $T$), to explain the acceleration of the universe~\cite{Bengochea:2008gz}. Note that $f(R)$ gravity is a higher-order theory, while the field equations of $f(T)$ gravity still remain second order. Subsequently, $f(T)$ gravity has been widely investigated in cosmology. The cosmological perturbations in $f(T)$ gravity were investigated in Ref. \cite{Chen:2010va}. Gravitational waves in $f(T)$ gravity were investigated in Ref. \cite{Farrugia:2018gyz}. For more researches on $f(T)$ cosmology, see Refs.~\cite{Cai:2015emx,Ferraro:2006jd,Bamba:2012vg,Fiorini:2013hva,Geng:2014nfa,Li:2018ixg}.

In 2012, the thick brane model in $f(T)$ gravity was firstly constructed in Ref.~\cite{Yang:2012hu}, the thick brane solutions were obtained, and the corresponding localization of fermions was also investigated. After that, using the superpotential method~\cite{Zhong:2016glr}, more thick brane solutions in $f(T)$ theory were obtained~\cite{Menezes:2014bta}. The tensor perturbations of the vielbein of $f(T)$ brane and the stability of this system were studied in Ref.~\cite{Guo:2015qbt}. It was found that the zero mode of the perturbation is localized on the brane. Then, in 2018, the braneworld model of $f(T)$ gravity with noncanonical scalar matter field (K-fields) was studied in Ref.~\cite{Wang:2018jsw}. More $f(T)$-brane related studies can be found in Refs.~\cite{DavoodSadatian:2018fss,Bamba:2013fta,Atazadeh:2014joa,Correa:2015qma,Guo:2018tpo}. In this paper we investigate the effects of torsion on thick branes structure and the resonance spectrum of KK gravitons. Based on Ref.~\cite{Guo:2015qbt}, we would like to study the gravitational resonances of $f(T)$-brane. The fluctuation equation of $f(T)$ gravity was given in Ref.~\cite{Guo:2015qbt}. Note that the dynamical variables are the vielbein fields, but we will focus on the gravitational resonances described by the metric, for which the relations between the perturbed vielbein and perturbed metric should be well defined. Firstly, the perturbed vielbein can be defined as follows
\begin{eqnarray}
	{e^A}_M=\left(
	\begin{array}{cc}
		e^{A(y)}({\delta^a}_\mu +{h^a}_\mu) & 0\\
		0 & 1\\
	\end{array}
	\right).
\end{eqnarray}
Using the relation between the metric and the vielbein
\begin{equation}
	g_{MN}=e^{A}_{{~M}}e^{B}_{~N}\eta_{AB},
\end{equation}
we can get
\begin{eqnarray}
	\gamma_{\mu\nu}\!\!&=&\!\!({\delta^a}_\mu {h^b}_\nu+{\delta^b}_\nu {h^a}_\mu)\eta_{ab}.
\end{eqnarray}
Then the tensor perturbation of the background metric is~\cite{Guo:2015qbt}
\begin{eqnarray}
	g_{MN}=\left(
	\begin{array}{cc}
		e^{2A(y)}(\eta_{\mu\nu}+\gamma_{\mu\nu}) & 0\\
		0 & 1\\
	\end{array}
	\right).
\end{eqnarray}
The equation of motion for the tensor perturbation can be gotten as follows
\begin{equation}
	\left(\partial_z^2+2H\partial_z+\Box^{(4)}\right)\gamma_{\mu\nu}=0, \label{perturbationEq2}
\end{equation}
where
\begin{eqnarray}
	H&=& \frac{3}{2}\partial_z A
	+12e^{-2A}\left(\left(\partial_z A \right)^3
	-\partial_z^2 A\partial_z A
	\right)
	\frac{f_{TT}}{f_T}   \label{H}           .
\end{eqnarray}
Obviously, the resonance spectrum of KK gravitons is closely related to the form of the function $f(T)$.
The spacetime torsion will cause the thick brane to split, making the thick brane appear more abundant internal structures. Therefore, more abundant resonance spectrum of KK gravitons may appear. In addition, it is also possible to reflect the structure of the extra dimension by studying the resonance spectrum of KK gravitons.

In this paper, we will construct some new $f(T)$-brane solutions. It will be shown that these $f(T)$-brane solutions  are stable under the transverse-traceless tensor perturbation. Based on these solutions, we will study the effects of different parameters on the effective potential and the gravitational resonances. We will also give the region where the resonance exists in the parameter space. Besides, the effects of the torsion will also be studied by comparing different kinds of $f(T)$. More importantly, we will investigate the possibility that the first resonance to be a candidate for dark matter. For the warp factor that we choose, the first resonance can not be a candidate for dark matter for $f(T)=T$ and $f(T)=T+\alpha T^{2}$. For $f(T)=T$, the long-lived resonance requires a very large thickness of the brane, which is inconsistent with the gravitational experiment. The form of $f(T)=T+\alpha T^{2}$ has the similar result with $f(T)=T$. But for the case of $f(T)=-T_{0}\left(e^{-\frac{T}{T_{0}}}-1\right)$,
the effect of torsion on both the effective potential and the lifetime of the first resonance can be large enough with suitable choice of the parameter $T_0$. In this case, we do not need a thick brane with a very large thickness to make the lifetime of the first resonance long enough to be a candidate of dark matter.

The organization of this paper is as follows. In Sec.~\ref{BRANE WORLD MODEL}, we will review $f(T)$-brane and its tensor perturbations~\cite{Guo:2015qbt}, and get the zero mode normalization condition. In Sec.~\ref{Gravity resonances}, we will construct some thick $f(T)$-brane models. Then we will investigate the possibility of the first resonance in these models to be a candidate for dark matter. Finally, in Sec.~\ref{Conclusion}, we come to the conclusions and discussions.

\section{BRANE WORLD MODEL IN $f(T)$ GRAVITY}
\label{BRANE WORLD MODEL}

Firstly, we give a brief review of the teleparallel gravity. This gravity theory was proposed by Einstein as an attempt of a unified theory of electromagnetism and gravity
on the mathematical structure of distant parallelism. We usually use the veilbein fields $e_{A}(x^{M})$ instead of the metric field $g_{MN}$ to describe the dynamics and structure of the spacetime. This is done in the tangent space associated with a spacetime point in the manifold, instead of the coordinate basis. These vielbein fields form an orthonormal basis of the tangent space at each point in the manifold with spacetime coordinates $x^{M}$. The relation between the spacetime metric and the veilbein fields is
given by
\begin{equation}
	g_{MN}=e^{A}_{{~M}}e^{B}_{~N}\eta_{AB},
\end{equation}
where $\eta_{AB}=\text{diag}(-1,1,1,1,1)$ is the Minkowski metric (in this paper, we focus on five dimensions). In this paper, capital Latin indices $A,B,C,\cdots$= 0,1,2,3,5 label tangent
space coordinates, while $M,N,O,\cdots= 0,1,2,3,5$ label spacetime ones. In this gravity theory, the spacetime is characterized by a curvature-free linear connection, i.e., the Weitzenb\"{o}ck connection $\tilde\Gamma_{MN}^{P}$ which is defined in terms of the vielbein fields:
\begin{equation}
	\tilde\Gamma_{~MN}^{P}\equiv e_{A}^{~P}\partial_{N}e_{~M}^{A}=-e_{~M}^{A}\partial_{N}e_{A}^{~P}.
\end{equation}
We use the Weitzenb\"{o}ck connection rather than the Levi-Civita connection $\Gamma_{MN}^{P}$ to define the associated tensors.
The torsion tensor is constructed from the Weitzenb\"{o}ck connection as
\begin{equation}
	T^{P}_{~MN}=\tilde\Gamma_{~NM}^{P}-\tilde\Gamma_{~MN}^{P}=e_{A}^{~P}(\partial_{N}e_{~M}^{A}-\partial_{N}e_{~M}^{A}).
\end{equation}
The difference between the Weitzenb\"{o}ck connection and the Levi-Civita connection is given by
\begin{eqnarray}
	K^{P}_{~MN}&=&\tilde\Gamma_{~MN}^{P}-\Gamma_{~MN}^{P}\nn\\
	&=&\frac{1}{2}(T_{M~N}^{~~P}+T_{N~M}^{~~P}-T_{~MN}^{P}).
\end{eqnarray}
It is useful to define another tensor $S_{P}^{~MN}$
\begin{equation}
	S_{P}^{~MN}\equiv\frac{1}{2}
	\left({K^{MN}}_{P}
	-{\delta^{N}_{P}{T^{QM}}_{Q}}
	+\delta^{M}_{P}{T^{QN}}_{Q}\right). \label{SPMN}
\end{equation}
So the torsion scalar $T$ is given by
\begin{equation}
	T\equiv S_{P}^{~~MN}T_{~~MN}^{P}.
\end{equation}
The Lagrangian of the teleparallel gravity can be written as
\begin{equation}
	L_T=-\frac{M_5^3}{4}e\, T, \label{Lagrangian}
\end{equation}
where $e$ is the determination of the vielbein $e_{A}(x^{M})$ and $M_5$ is the five-dimensional mass scale, which is set to $M_5=1$ in this paper. It is known that the teleparallel
gravity is equivalent to general relativity and hence is also called as the teleparallel equivalent of general relativity since $R=-T-2\nabla^{M}T^{N}_{~~MN}$.

In $f(T)$ gravity the torsion scalar is replaced with $f(T)$, a function of \emph{T}. When $f(T)=T$, it goes back to teleparallel gravity, and hence is equivalent to GR. In a five-dimensional $f(T)$ gravity, the action is given by
\begin{equation}
	S=-\frac{1}{4}\int d^5x~ e~ f(T)+\int d^5x \mathcal{L}_M,
	\label{action}
\end{equation}
where $\mathcal{L}_M$ denotes the Langrangian density of the matter. After varying the action with respect to the vielbein, we can get the field equation:
\begin{eqnarray}
	&&e^{-1}f_T g_{NP}\partial_Q \left(e\,S_{M}^{~~PQ}\right)+f_{TT}S_{MN}^{~~~~Q}\partial_Q  T \nonumber\\ &&-{f}_{T}\tilde\Gamma^P_{~~QM}S_{\!PN}^{~~~~Q}+\frac{1}{4}g_{MN}f(T)=\mathcal{T}_{MN},
	\label{field equation}
\end{eqnarray}
where $f_T\equiv\frac{df(T)}{dT}$, $f_{TT}\equiv\frac{d^2 f(T)}{d T^2}$, and $\mathcal{T}_{MN}$ is the  energy-momentum tensor of the matter field.

Now, we would like to consider thick brane models in $f(T)$ gravity theory. The metric of the flat brane with codimension one is given by
\begin{equation}
	ds^2=e^{2A(y)}\eta_{\mu\nu}dx^\mu dx^\nu+dy^2,
	\label{metric}
\end{equation}
where $\eta_{\mu\nu}=\text{diag}(-1,1,1,1)$ is the four-dimensional Minkowski metric and $e^{2A(y)}$ is the wrap factor.
The bulk vielbein is
\begin{equation}
	e^{A}_{~M}=\text{diag}(e^{A(y)},e^{A(y)},e^{A(y)},e^{A(y)},1),    \label{vielbein}
\end{equation}
and the torsion scalar is
\begin{equation}
	T=-12 A'(y)^2.    \label{T}
\end{equation}
We choose the matter Lagrangian density as
\begin{equation}
	\mathcal{L}_M=e\big(-\frac{1}{2}\partial^M\phi~\partial_M\phi-V(\phi)\big),
	\label{energy momentum}
\end{equation}
where $\phi$ is the background scalar field only depending on the extra dimension $y$, and $V(\phi)$ is the potential of the scalar field $\phi$. For such setup, the explicit equations
of motion are given by
\begin{eqnarray}
	&&6A'^2 f_T+\frac{1}{4}f=
	-V+\frac{1}{2}\phi'^2,  \label{EoMs1}\\
	&&\frac{1}{4}f+\left(\frac{3}{2}A{''}+6A'^2\right)f_T
	-36A'^2A{''}f_{TT}=-V-\frac{1}{2}\phi'^2,  \label{EoMs2}\nn\\
	\\
	&&\phi{''}+4A'\phi'=\frac{dV}{d\phi}.  \label{EoMsphi}
\end{eqnarray}
It can be shown that only two of the above three equations are independent. So we need to give two of the four variables to solve these equations. From  Eq.~(\ref{EoMs1}) and Eq.~(\ref{EoMs2}), we can obtain that
\begin{eqnarray}
	V&=&-\frac{1}{4}f-\left(\frac{3}{4}A''+6A'^{2}\right)f_{T}+18A'^{2}A''f_{TT},\label{V field}\\
	\phi'^{2}&=&36A'^{2}A''f_{TT}-\frac{3}{2}A''f_{T}.\label{phi field}
\end{eqnarray}

Next, we consider the linear transverse-traceless tensor perturbation of the metric corresponding to the vielbein, which was investigated firstly in Ref.~\cite{Guo:2015qbt}. The perturbed vielbein can be written as
\begin{eqnarray}
	{e^A}_M=\left(
	\begin{array}{cc}
		e^{A(y)}({\delta^a}_\mu +{h^a}_\mu) & 0\\
		0 & 1\\
	\end{array}
	\right),
\end{eqnarray}
where the Latin letters $a,b,\cdots$ denote the tangent space coordinates on the brane, and the Greek letters $\mu,\nu \cdots$ denote the spacetime coordinates on the brane. And then, the tensor perturbation of the background metric can be written as
\begin{eqnarray}
	g_{MN}=\left(
	\begin{array}{cc}
		e^{2A(y)}(\eta_{\mu\nu}+\gamma_{\mu\nu}) & 0\\
		0 & 1\\
	\end{array}
	\right),
\end{eqnarray}
where
\begin{eqnarray}
	\gamma_{\mu\nu}\!\!&=&\!\!({\delta^a}_\mu {h^b}_\nu+{\delta^b}_\nu {h^a}_\mu)\eta_{ab},
\end{eqnarray}
satisfies the transverse-traceless condition
\begin{eqnarray}
	\partial_\mu\gamma^{\mu\nu}=0=\eta^{\mu\nu}\gamma_{\mu\nu}.
\end{eqnarray}
Considering the above conditions, we obtain the main equation of the tensor perturbation~\cite{Guo:2015qbt}:
\begin{eqnarray}
	&& \left(e^{-2A}\Box^{(4)}\gamma_{\mu\nu}+\gamma''_{\mu\nu}+4A'\gamma'_{\mu\nu}\right)
	f_T \nn \\
	&&-24A'A''\gamma'_{\mu\nu}f_{TT}=0,
	\label{mainequation}
\end{eqnarray}
where $\Box^{(4)}=\eta^{\mu\nu}\partial_{\mu}\partial_{\nu}$. With the coordinate transformation
\begin{equation}
	dz=e^{-A}dy,
\end{equation}
Eq.~(\ref{mainequation}) becomes
\begin{equation}
	\left(\partial_z^2+2H\partial_z+\Box^{(4)}\right)\gamma_{\mu\nu}=0, \label{perturbationEq2}
\end{equation}
where
\begin{eqnarray}
	H&=& \frac{3}{2}\partial_z A
	+12e^{-2A}\left(\left(\partial_z A \right)^3
	-\partial_z^2 A\partial_z A
	\right)
	\frac{f_{TT}}{f_T}   \label{H}           .
\end{eqnarray}
Now, we introduce the KK decomposition
\begin{equation}
	\gamma_{\mu\nu}(x^\rho,z)=\epsilon_{\mu\nu}(x^\rho) F(z) \psi(z), \label{KK decomposition}
\end{equation}
where
\begin{eqnarray}
	F(z) &=& e^{-\frac{3}{2}A(z)+\int{K(z)dz}},\\
	K(z)&=&12e^{-2A}\Big(\partial_z^2 A\partial_z A
	-\left(\partial_z A \right)^3
	\Big)
	\frac{f_{TT}}{f_T}.
\end{eqnarray}
Substituting Eq.~(\ref{KK decomposition}) into Eq.~(\ref{perturbationEq2}), we get two equations: the Klein-Gordon equation for the four-dimensional KK gravitons $\epsilon_{\mu\nu}$:
\begin{eqnarray}
	\left(\Box^{(4)}+m^2\right)\epsilon_{\mu\nu}(x^\rho)= 0, \label{EoMepsilonMuNu}
\end{eqnarray}
and the Schr\"odinger-like equation for the extra-dimensional profile:
\begin{eqnarray}
	\left(-\partial_z^2+U(z)\right)\psi = m^2\psi, \label{SchrodingerEquation}
\end{eqnarray}
where $m$ is the mass of the KK graviton and the effective potential is given by~\cite{Guo:2015qbt}
\begin{eqnarray}
	U(z)=H^2+\partial_z H. \label{EffectivePotential}
\end{eqnarray}
The Schr\"odinger-like equation~(\ref{SchrodingerEquation}) can be factorized as
\begin{equation}
	\big(\partial_z+H\big)\big(-\partial_z+H\big)\psi=m^2\psi,
\end{equation}
which ensures that the eigenvalues $m^{2}$ are non-negative, so there is no tensor tachyon mode with $m^2<0$. That is to say any brane solution of $f(T)$ gravity theory is stable under the transverse-traceless tensor perturbation.
The solution of the graviton zero mode (the four-dimensional massless graviton) is
\begin{eqnarray}
	\psi_0=N_0e^{\frac{3}{2}A-\int K(z)dz},\label{zeromodefunc}
\end{eqnarray}
where $N_0$ is the normalization coefficient. Note that, in order to recover the four-dimensional Newtonian potential on the brane, the zero mode of graviton should satisfy the following normalization condition
\begin{equation}
	\int dz \;\psi_{0}^2(z) < \infty. \label{NormalizationCondition}
\end{equation}

\section{Gravitational resonances in various $f(T)$-brane models}
\label{Gravity resonances}

In this section, we will give some solutions of braneworld and investigate the gravitational resonances in thick $f(T)$-braneworld models. Because $f(T)$ is an arbitrary function of the torsion scalar $T$, so different functional forms can give different solutions.

In this paper, we consider the following warp factor
\begin{eqnarray}
	A(y)=\ln\Big[\tanh\big(k(y+b)\big)-\tanh \big(k(y-b)\big)\Big].\label{warp factor}
\end{eqnarray}
Here the parameter $k$ has mass dimension one. The parameter $b$ has length dimension one and denotes the distance of two sub-branes. For convenience, we define the dimensionless scaled distance $\bar{b}=kb$.  The shape of the warp factor is shown in Fig.~\ref{figwarp factor}, from which we can see that there is a platform near $y=0$ for large $\bar b$. When $y\rightarrow \pm\infty$, $A(y)\rightarrow-2k|y|$, so the spacetime is asymptotically AdS$_{5}$.

\subsection{Model 1: $f(T)=T$}
\begin{figure}
	\centering
	\subfigure[~The warp factor (\ref{warp factor})]{\label{figwarp factor}
		\includegraphics[width=0.22\textwidth]{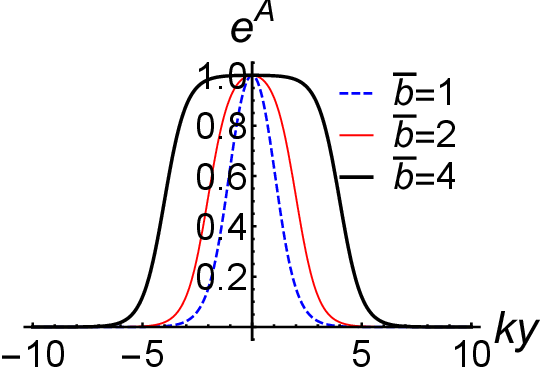}}
	\subfigure[~The scalar field (\ref{grscalar}) ]{\label{figgrscalar}
		\includegraphics[width=0.22\textwidth]{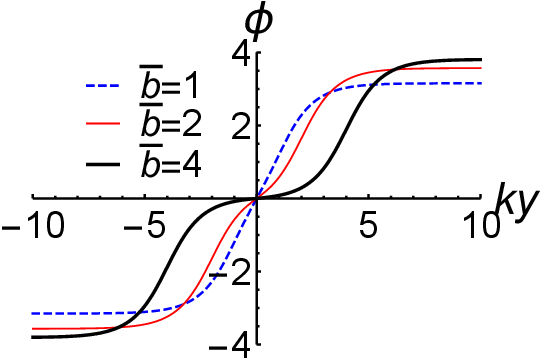}}
	\subfigure[~The scalar potential $V(\phi)$ ]{\label{1scalarpotential}
		\includegraphics[width=0.22\textwidth]{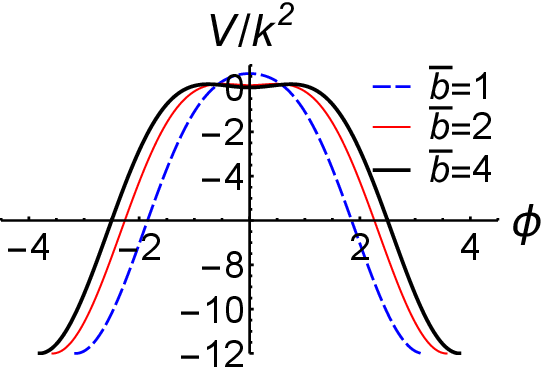}}
	\caption{Plots of the warp factor (\ref{warp factor}) and the scalar field (\ref{grscalar}), and the scalar potential $V(\phi)$.}\label{figwarpfactorandscalar}
\end{figure}

We first consider $f(T)=T$, which is equivalent to GR. From Eqs.~(\ref{V field}) and~(\ref{phi field}) we get the solution
\begin{eqnarray}
	\phi(y)&=&-i \sqrt{3} \text{sech}(b k)\Big[\cosh (2 b k) \text{F}\big(i k y;\tanh ^2(b k)+1\big)\nn\\
	&&-2 \sinh ^2(b k) {\rm \Pi} \big(\text{sech}^2(b k);i k y;\tanh ^2(b k)+1\big)\Big],\label{grscalar}\nn\\ \\
	V(y)&=&\frac{3}{4} k^2 \Big[-4 \big(\tanh (k (y-b))+\tanh (k (b+y))\big)^2\nn\\
	&&+\text{sech}^2(k (y-b))+\text{sech}^2(k (b+y))\Big],
\end{eqnarray}
where $\text{F}(y;q)$ and ${\rm\Pi}(y;q;p)$ are the first and third kind elliptic integrals, respectively. Plots of the scalar field are shown in Fig.~\ref{figgrscalar}. It can be seen that the scalar field has the configuration of a single kink for small $\bar{b}$. With the increase of $\bar{b}$, the single kink becomes a double kink, and the value of $|\phi(\pm\infty)|$ increases accordingly. In particular, when $\bar{b}\rightarrow0$, $\phi(\pm\infty) \rightarrow \pm\frac{\sqrt{3}\pi}{2}\simeq\pm 2.72$;  when $\bar{b}\rightarrow \infty$, $\phi(\pm\infty) \simeq \pm3.85$.  In general, the appearance of the double kink means that the brane splits into two sub-branes. The distance between the two sub-branes is $b$. Because the solution of the scalar field is more complicated, we have not inversely solved the expression of $V(\phi)$. Numerically, we plot the shape of the scalar potential $V(\phi)$ in Fig.~\ref{1scalarpotential}. It can be seen that as $\bar{b}$ increases, $V(\phi)$ splits at $\phi= 0$.

\begin{figure}
	\centering
	\subfigure[~The effective potential (\ref{GReffectiveP})]{\label{figGReffectiveP}
		\includegraphics[width=0.22\textwidth]{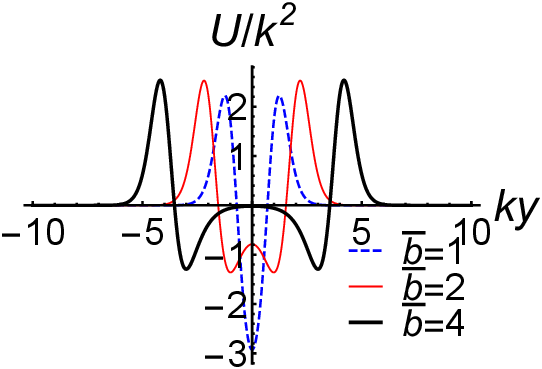}}
	\subfigure[~$\bar{b}=1$]{\label{grP1}
		\includegraphics[width=0.22\textwidth]{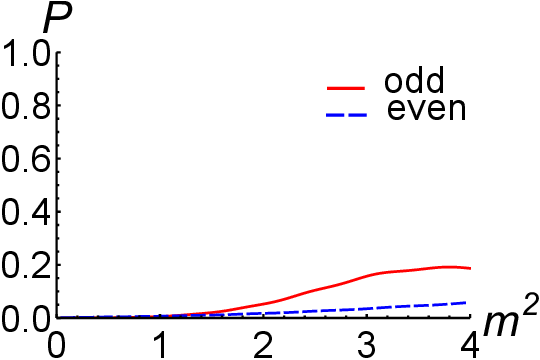}}
	\subfigure[~$\bar{b}=2$]{\label{grP2}
		\includegraphics[width=0.22\textwidth]{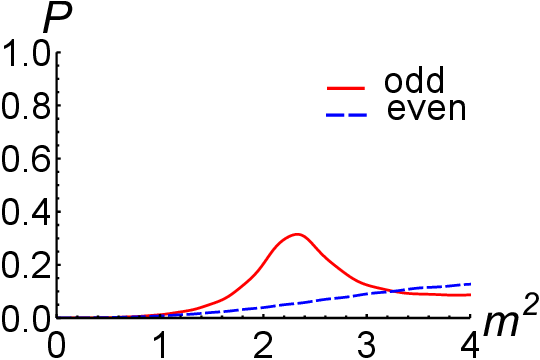}}
	\subfigure[~$\bar{b}=4$]{\label{grP3}
		\includegraphics[width=0.22\textwidth]{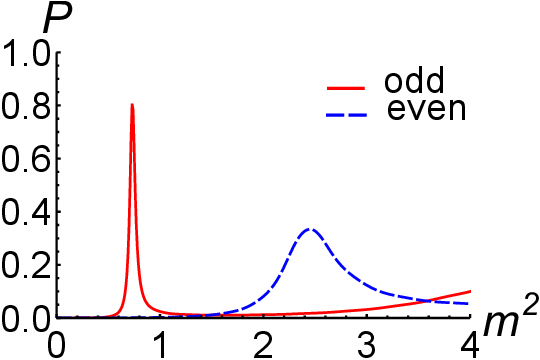}}
	\caption{The influence of the parameter $\bar{b}$ on the effective potential (\ref{GReffectiveP}) and the relative probability $P$ of $f(T)=T$ for the odd-parity (red lines) and even-parity (blue dashed lines) massive KK modes.}\label{model1P}
\end{figure}

The effective potential (\ref{EffectivePotential}) in the coordinate $y$ is given by
\begin{eqnarray}
	U(z(y))&=&-\frac{3}{8}k^2 \text{sech}^2\big(k (b-y)\big) \text{sech}^2\big(k (b+y)\big)\nn\\
	&&\times \Big(\tanh \big(k (b-y)\big)
	+\tanh \big(k (b+y)\big)\Big)^{2} \nn\\
	&&\times\Big(-5\cosh (4 k y)+2 \cosh \big(2 k (b-y)\big)\nn\\
	&&+2 \cosh \big(2 k (b+y)\big)+9\Big)
	\label{GReffectiveP}.
\end{eqnarray}
We plot the shape of the above effective potential in Fig.~\ref{figGReffectiveP}.
It can be seen that the depth of the potential well decreases with the parameter $\bar{b}$, and the height of the potential barrier increases with $\bar{b}$. Both changes become smaller with $\bar{b}$. On the other hand, it can be seen that the width of the effective potential increases with $\bar{b}$. With the increase of $\bar{b}$, the potential splits from one well to two wells. We will see later that the appearance of the two-well structure will result in gravitational resonances.

\begin{figure}
	\centering
	\subfigure[~$\bar{m}_1(\bar{b})$]{\label{figgrbm}
		\includegraphics[width=0.22\textwidth]{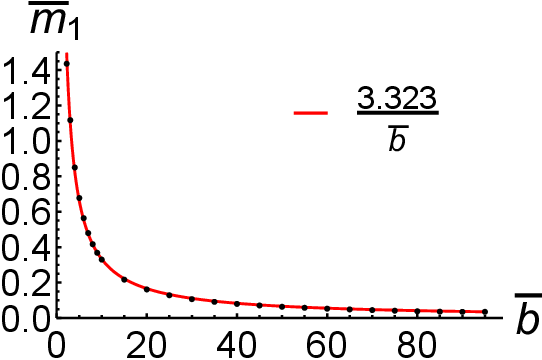}}
	\subfigure[~$\bar{\tau}_1(\bar{b})$]{\label{giggrblife}
		\includegraphics[width=0.22\textwidth]{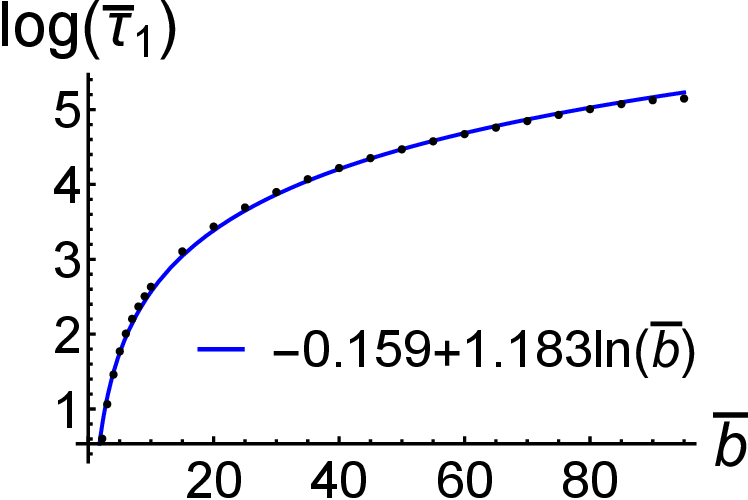}}
	\caption{The relations of the scaled mass $\bar{m}_{1}$ and the scaled lifetime $\bar{\tau}_{1}$ of the first resonance with the parameter $\bar{b}$ for $f(T)=T$. The dots are calculated values while the red and blue lines are fit functions.}\label{model1mt}
\end{figure}

Next, we will investigate the gravitational resonances in this model. Inspired by the investigation of Ref.~\cite{Liu:2008pi}, Almeida $et~al$ firstly proposed a method to find the fermion resonances by using large peaks in the distribution of  the normalized squared wavefunction~\cite{Almeida:2009jc}. But this method is only applicable to even functions, and it is no longer valid when the solutions are odd. In order to find all resonances, Liu $et~al$ proposed another method by defining the relative probability~\cite{Liu:2009ve}
\begin{eqnarray}
	P(m^{2})=\frac{\int^{z_{b}}_{-z_{b}}|\psi(z)|^{2}dz}
	{\int^{z_{max}}_{-z_{max}}|\psi(z)|^{2}dz},\label{relative probability}
\end{eqnarray}
where $\psi(z)$ is the solution of Eq.~(\ref{SchrodingerEquation}), $z_b$ is approximately the width of the brane, and $z_\mathrm{max}=10z_b$. Here $|\psi(z)|^{2}$ can be explained as the probability density. If the relative probability $P(m^2)$ has a peak around $m=m_n$ and this peak has a full width at half maximum, we can say that there exists a resonant mode with mass $m_n$. Since the potentials considered in this paper are symmetric, the wave functions are either
even or odd. Hence, we can use the following boundary conditions to solve the differential equation (\ref{SchrodingerEquation}) numerically:
\begin{subequations}
	\begin{eqnarray}
		\label{even}
		\psi_{\rm{even}}(0)\!\!&=&\!\!1, ~~~\partial_{z}\psi_{\rm{even}}(0)=0;\\
		\label{odd}
		\psi_{\rm{odd}}(0)\!\!&=&\!\!0, ~~~~\partial_{z}\psi_{\rm{odd}}(0)=1,
	\end{eqnarray}\label{EvenOddConditions}
\end{subequations}
where $\psi_{\rm{even}}$ and $\psi_{\rm{odd}}$ denote the even and odd modes of $\psi(z)$, respectively. Then, substituting the effective potential (\ref{GReffectiveP}) into Eq.~(\ref{SchrodingerEquation}) we can obtain the solution of the
extra-dimensional profile $\psi(z)$ with mass $m$ and hence the relative probability $P(m^2)$. According to the supersymmetric quantum mechanics, the supersymmetric partner potentials will share the same spectrum of massive excited states. So, we can judge
whether there are resonances by analyzing the shape of the supersymmetric partner potential. The dual potential corresponding to the effective potential (\ref{EffectivePotential}) is $U^{(\text{dual})}(z)=H^{2}-\partial_{z}H$. If there is no well or quasi-well in the dual potential, then
there is no resonances \cite{Zhang:2016ksq}. Thus, for $f(T)=T$, only for $\bar{b}>1$, there might exist resonances. Just as said before, the width of the potential barrier increases with $\bar{b}$, which indicates there are more resonances for larger $\bar{b}$, which can be seen from Figs.~\ref{grP1}, \ref{grP2}, \ref{grP3}. Furthermore, we can obtain the corresponding lifetime $\tau$ of the gravitational resonances by the width $\Gamma$ at half maximum of the peak, i.e., $\tau=\frac{1}{\Gamma}$. For convenience, we define the dimensionless scaled mass $\bar{m}_{n}=m_{n}/k$ and the dimensionless scaled lifetime $\bar{\tau}=k\tau$. The relations of the scaled mass $\bar{m}_{1}$ and the scaled lifetime $\bar{\tau}_1$ of the first resonance with the parameter $\bar{b}$ are shown in Fig.~\ref{model1mt}. It can be seen that, the scaled mass $\bar{m}_{1}$ of the first resonance decreases with $\bar{b}$, while the scaled lifetime $\bar{\tau}_1$ of the first resonance increases with $\bar{b}$. This behaviour means that if the parameter $\bar{b}$ is large enough, the lifetime of the first resonance with very light mass can be long enough as the age of our universe. So, we can consider such first gravitational resonance as one of the candidates for dark matter~\cite{Sui:2020fty}.

\begin{figure}
	\centering
	\subfigure[~$k-\bar{b}$]{\label{figgrlimitk-b}
		\includegraphics[width=0.23\textwidth]{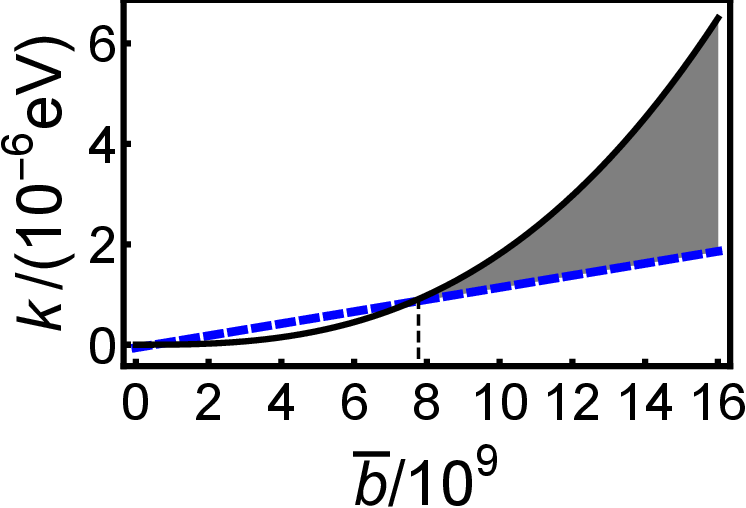}}
	\subfigure[~$m_{1}-\bar{b}$]{\label{figgrlimitm-b}
		\includegraphics[width=0.23\textwidth]{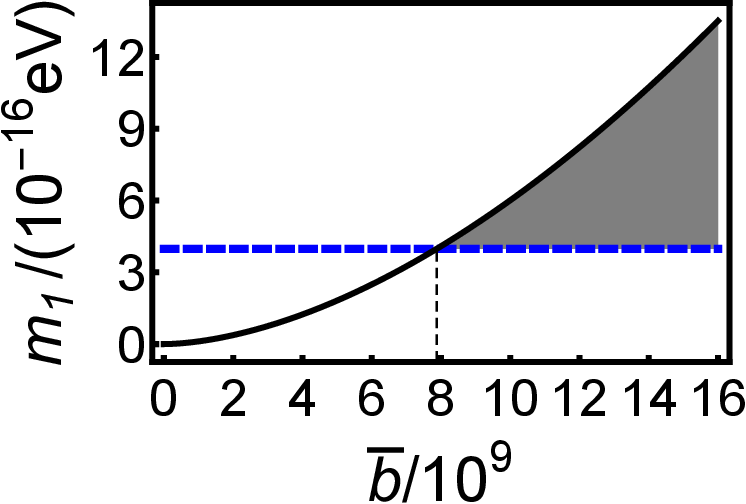}}
	\caption{The limit range of the parameter $k$ and the corresponding range of the first resonance mass $m_{1}$. The shadow region of the left panel is the limit range of the parameter $k$ and the shadow region of the right panel is the corresponding range of the first resonance mass $m_{1}$. The black solid lines are the limits that the lifetime of the first resonance should be longer than the age of the universe and the blue dashed lines are the limits from the five-dimensional fundamental scale $M_5$ should be greater than 13~TeV.}\label{model1limit}
\end{figure}

The scaled mass $\bar{m}_1$ and the scaled lifetime $\bar{\tau}_{1}$ can be fitted as two functions of the parameter $\bar b$, and the fit functions are given by
\begin{eqnarray}
	\bar{m}_{1}&=&\frac{3.323}{\bar{b}}\label{fitm1},\\
	\log(\bar{\tau}_{1})&=&-0.159+1.183\ln(\bar b)\label{fitt1}.
\end{eqnarray}
In the brane world theory considered in this paper, the relation between the effective four-dimensional Planck scale $M_{\text{Pl}}$ and the fundamental five-dimensional scale $M_{5}$ is given by:
\begin{equation}
	M_{\text{Pl}}^2=M_{5}^3\int_{-\infty}^{\infty} dy e^{2A(y)}f_{T},  \label{conditionMall}
\end{equation}
for $f(T)=T$, Eq.~(\ref{conditionMall}) becomes
\begin{equation}
	M_{\text{Pl}}^2=\frac{8\bar{b}\coth(2\bar{b})-4}{k}M_{5}^3.\label{conditionM}
\end{equation}
According to the recent experiment of the Large Hadron Collider (LHC),  { the collision energy is 13 TeV} and the result shows that the quantum effect of gravity can be ignored. Theoretically, the quantum effect of gravity will appear if the energy scale is larger than the  five-dimensional fundamental scale $M_{5}$. Thus, from the condition $M_5>13$ TeV and Eq.~(\ref{conditionM}), we can give the constraint on the parameter $k$ in the Natural System of Units:
\begin{equation}\label{limitM}
	k > \big(12\bar{b}\coth(2\bar{b})-6\big)\times{10^{-17}}~\text{eV}.
\end{equation}
On the other hand, it is known that the age of our universe { is of about 13.8} billion years, i.e.,  $4.35\times10^{17}~\text{s}$. So, if we consider the first resonance as one of candidates for dark matter, its lifetime should be { larger than the age of universe}, i.e.,  $\tau_1 \gtrsim 4.35 \times 10^{17}~\text{s}$, or in the Natural System of Units,
\begin{equation}\label{conditiontau}
	\tau_1=\bar{\tau}_1/k \gtrsim 6.6\times10^{32}~\text{eV}^{-1}.
\end{equation}
Thus, the restriction of the parameter $k$ can be expressed as
\begin{equation}
	k \lesssim 1.5 \times{10^{-33}} \bar{\tau}_1~\text{eV}\simeq  \bar{b}^{2.724}\times ~10^{-33}~\text{eV}.\label{limitkb}
\end{equation}
By combining the fit function \eqref{fitm1} and the two conditions \eqref{limitM}, \eqref{limitkb}, the restricted expressions of the mass of the first resonance $m_{1}$ with the combination parameter $\bar b$ can be obtained
\begin{eqnarray}
	m_1&>&\frac{3.323}{\bar{b}}\big(12\bar{b}\coth(2\bar{b})-6\big)\times{10^{-17}}~\text{eV}\label{conditionm1},\\
	m_1&\lesssim&3.456\bar{b}^{1.724}~\times{10^{-33}}~\text{eV}\label{conditionm2}.
\end{eqnarray}
The shadow regions of Fig.~\ref{model1limit} show the available ranges of the parameters $k$ and $m_{1}$, respectively. From Fig.~\ref{figgrlimitk-b}, we can see that only if
\begin{eqnarray}
	\bar{b}>7.9\times10^{9},\label{barblimit}
\end{eqnarray}
the two restricted conditions \eqref{limitM} and \eqref{limitkb} of $k$ could be satisfied, which means that the parameter $\bar{b}$ has a lower bound. And the corresponding constraint of the parameter $k$ is $k\gtrsim9.5\times{10^{-7}}~\text{eV}$. From Fig.~\ref{figgrlimitm-b}, we can see that the first resonance mass $m_{1}$ has a lower bound, i.e., $m_{1}\gtrsim 4\times 10^{-16}~\text{eV}$. But there are some problems here:

(1) In any realistic brane scenario, the matter fields on the brane will cover the entire thickness of the brane along the extra dimension. On the other hand, such a large $\bar{b}$ value means that the thickness of the brane is also very large. Within our constraints, $k$ is about $10^{-6}$eV and $\bar{b}$ is about $10^{10}$. The corresponding size of $b=\frac{\bar{b}}{k}$ is about $10^{11}$cm, which will cause the effective four-dimensional gravitational potential deviates from the squared inverse law at a very large distance.
According to the method of Ref.~\cite{Csaki:2000fc}, the correction term to the Newtonian potential from all massive gravitons in this model is
\begin{eqnarray}
	\Delta V(r)&=&G_N \frac{M}{r}\left(\int_{-\infty}^{\infty}dy e^{2A}\right)\int_{0}^{\infty}dm \frac{me^{-mr}}{k}\nn\\
	&=&G_N \frac{M}{r}\frac{8\bar{b}\coth (2\bar{b})-4}{k^2 r^2}.
\end{eqnarray}
We can see that the four-dimensional effective Newtonian potential becomes
\begin{eqnarray}
	V(r)=G_N \frac{M}{r}\left(1+\frac{8\bar{b}\coth (2\bar{b})-4}{k^2 r^2}\right).
\end{eqnarray}
So when $\bar{b}=10^{10}$, the correction scale $r$ is about $10^{7}$cm. This is contradicted with the current experiments of gravitational inverse-square law.
In fact, Kiritsis et al~\cite{Kiritsis:2001bc} showed that in their thick brane model, when the thickness of the brane is large enough, the Newtonian potential of two points close to each other on the brane becomes five-dimensional.

(2) Note that, we use the relative probability method \eqref{relative probability} to obtain the resonance  and give the lifetime of the first KK resonance state at $\bar{b}\leq10^2$. Although we cannot guarantee that the fit functions \eqref{fitm1} and \eqref{fitt1} are still valid when $\bar{b}$ is very large, we do not need to care about this. Because, the number of the resonances will increase with the parameter $\bar{b}$, and the height of the effective potential is almost unchanged. That is to say, the width at half maximum of the first resonance will decrease with $\bar{b}$.  It will result in that the lifetime of the first resonance increases with $\bar{b}$.  Based on the above discussion, maybe the value of the equation \eqref{barblimit} is not accurate, but our result is still valid. That is to say, under TEGR and warp factor \eqref{warp factor}, the first resonance can not be the candidate of dark matter. Therefore, we consider $f(T)$ gravity and see the effect of torsion.


\subsection{Model 2: $f(T)=T+\alpha T^{2}$}
\begin{figure}
	\centering
	{\includegraphics[width=0.23\textwidth]{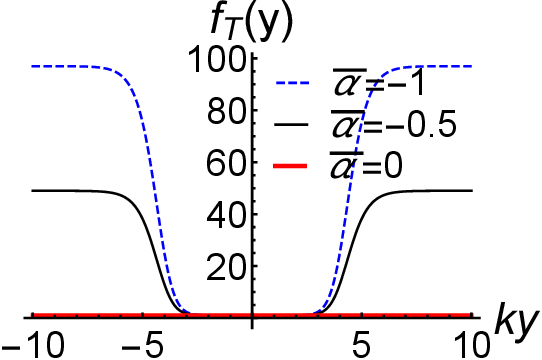}}
	\caption{Plots of the function $f_{T}$.}\label{ftcompare}
\end{figure}

Secondly, we consider the widely studied form of $f(T)$: $f(T) = T + \alpha T^2$ in brane world and cosmology, where the mass dimension of the parameter $\alpha$ is -2. For convenience, we define the dimensionless parameter $\bar\alpha = \alpha k^2$. The deviation of $f(T)$ gravity from GR can be denoted by the value of $f_{T}(y)$. As we can see from Fig.~\ref{ftcompare}, the larger $|\bar\alpha|$ is, the more deviation from GR, which corresponds to the case of $\bar{\alpha}=0$. In this model, Eq.~(\ref{phi field}) becomes
\begin{eqnarray}
	\phi'^{2}&=&\frac{3}{32} k^2 \text{sech}^4\big(k (y-b)\big) \text{sech}^4\big(k (b+y)\big)\nn\\
	&&\times\Big(\cosh \big(2 k (y-b)\big)+\cosh \big(2 k (b+y)\big)+2\Big) \nn\\
	&&\times\bigg[2 \cosh \big(2 k (y-b)\big)+ 2\cosh \big(2 k (b+y)\big)\nn\\
	&&+(1-288 \alpha  k^2)\cosh (4 k y) +\cosh (4 b k)  \nn\\
	&&+2+288 \alpha  k^2\bigg].\label{scalarfieldEq}
\end{eqnarray}
In order to ensure that the scalar field $\phi$ is real, the parameter $\alpha$ should satisfy $\alpha\leq \frac{1}{288k^{2}}$.

\begin{figure}
	\centering
	\subfigure[~$\bar{b}$=1]{\label{figmodel2scalar1}
		\includegraphics[width=0.22\textwidth]{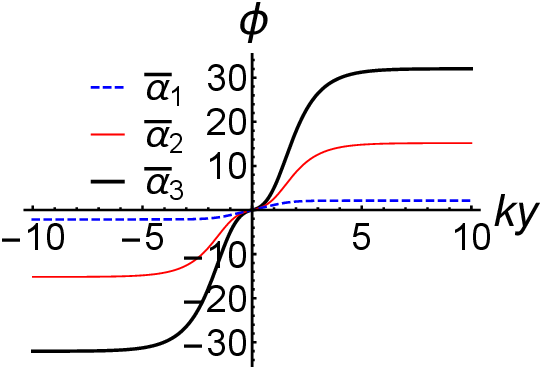}}
	\subfigure[~$\bar{b}$=4]{\label{figmodel2scalar2}
		\includegraphics[width=0.22\textwidth]{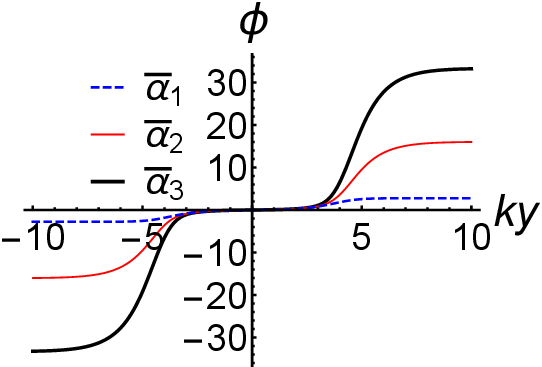}}
	\subfigure[~$\bar{b}$=1]{\label{figmodel2effectivep1}
		\includegraphics[width=0.22\textwidth]{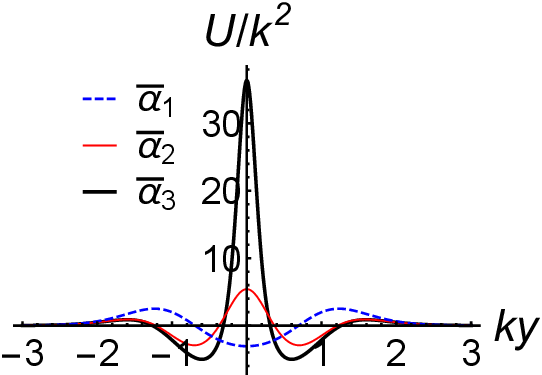}}
	\subfigure[~$\bar{b}$=4]{\label{figmodel2effectivep2}
		\includegraphics[width=0.22\textwidth]{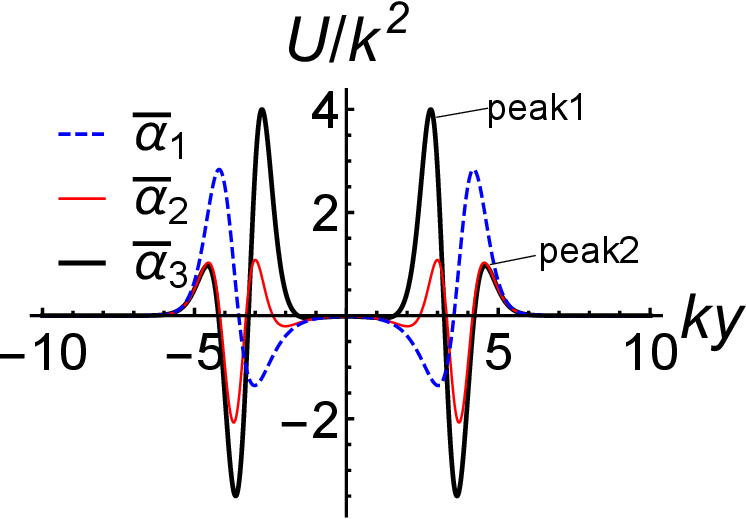}}
	\caption{Plots of the scalar field $\phi(y)$ and effective potential $U(y)$ for $f(T)=T+\frac{\bar{\alpha}}{k^{2}} T^{2}$. The parameter $\bar{b}$ is set to $\bar{b}=1,~4$. The parameter $\bar{\alpha}$ is set to $\bar{\alpha}_{1}=1/288 \approx 0.00347$ (blue dashed lines), $\bar{\alpha}_{2}=-0.212$ (red lines), and $\bar{\alpha}_{3}=-1$ (black thick lines). }\label{model2scalarandpotential}
\end{figure}

We can solve Eq.~(\ref{scalarfieldEq}) numerically and show the plots of the scalar field in Figs.~\ref{figmodel2scalar1} and~\ref{figmodel2scalar2}. It can be seen that, the shape of the scalar field depends on the values of $\bar{\alpha}$ and $\bar{b}$. The scalar field has the configuration of a single kink for small $|\bar{\alpha}|$ and $\bar{b}$. With the increase of $|\bar{\alpha}|$, the single kink becomes a double one, and the asymptotic value of $|\phi(\pm\infty)|$ increases accordingly. The influence of the parameter $\bar{b}$ on the scalar field is the same as the case of $f(T)=T$.

\begin{figure}
	\centering
	\subfigure[~$\bar{\alpha}=0.00347$]{\label{figmodel2P1}
		\includegraphics[width=0.22\textwidth]{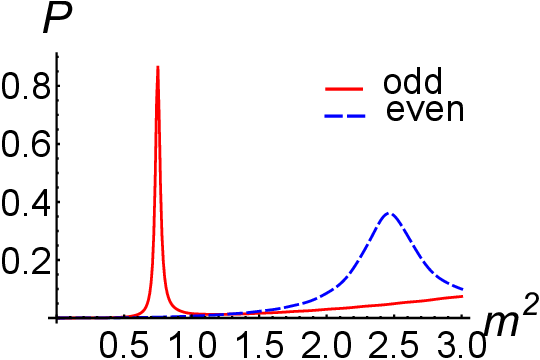}}
	\subfigure[~$\bar{\alpha}=-0.212$]{\label{figmodel2P2}
		\includegraphics[width=0.22\textwidth]{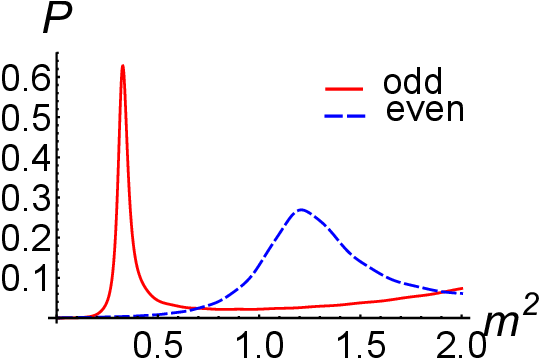}}
	\subfigure[~$\bar{\alpha}=-1$]{\label{figmodel2P3}
		\includegraphics[width=0.22\textwidth]{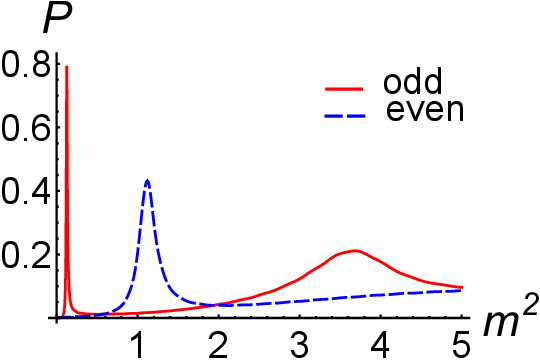}}
	\caption{The relative probabilities for different values of $\bar{\alpha}$ for $f(T)=T+\frac{\bar{\alpha}}{k^{2}} T^{2}$. The parameter is set to $\bar{b}=4$.}\label{model2P1}
\end{figure}

\begin{figure}
	\centering
	\subfigure[~$\bar{b}=1$]{\label{figmodel2p4}
		\includegraphics[width=0.22\textwidth]{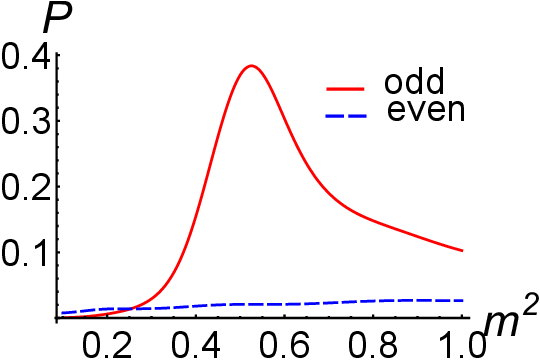}}
	\subfigure[~$\bar{b}=5$]{\label{figmodel2p5}
		\includegraphics[width=0.22\textwidth]{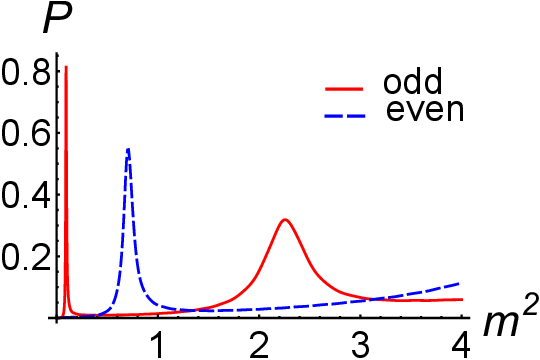}}
	\subfigure[~$\bar{b}=10$]{\label{figmodel2p5}
		\includegraphics[width=0.22\textwidth]{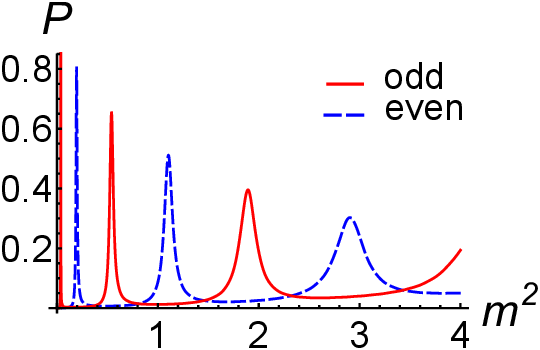}}
	\caption{The relative probability $P$ for different values of $\bar{b}$ for $f(T)=T+\frac{\bar{\alpha}}{k^{2}} T^{2}$. The parameter is set to $\bar{\alpha}=-1$.}\label{model2P2}
\end{figure}

Since the expression of the effective potential for $f(T)=T+\alpha T^{2}$ is complicated and tedious, we only show its plots in Figs.~\ref{figmodel2effectivep1} and \ref{figmodel2effectivep2}. From Fig.~\ref{figmodel2effectivep1} we can see that the depth of the effective potential well and the height of the potential barrier decrease with the parameter $\bar{\alpha}$. With the decrease of $\bar{\alpha}$, the potential splits from one well to two wells. From Fig.~\ref{figmodel2effectivep2} we can see that the effective potential has four peaks. As $-0.046<\bar{\alpha}\leq1/288 \approx0.00347$ for the case of $\bar{b}=4$, the four peaks will become two peaks. We label two of them as peak 1 and as peak 2 in Fig.~\ref{figmodel2effectivep2}. As $\bar{\alpha}$ increasing, the height of peak 1 decreases, but the height of peak 2 slowly increases. When $\bar{\alpha}=-0.212$, the heights of peak 1 and peak 2 are approximately the same. The relative probabilities $P(m^2)$ of the gravitational resonances for $\bar{b}=4, \bar{\alpha}=0.00347, -0.212$, and $-1$ are shown in Fig.~\ref{model2P1}.
In order to investigate the effects of the parameter $\bar{b}$, we fix $\bar{\alpha}=-1$. Plots of the relative probability for different values of $\bar{b}$ are shown in Fig.~\ref{model2P2}. We find that the number of resonances increases with $\bar{b}$, and the mass $m_1$ of the first resonance decreases with $\bar{b}$. This result is the same as $f(T)=T$. The relation between the number of the resonances and the parameters $\bar{\alpha}$ and $\bar{b}$ is shown in Fig.~\ref{model2alphab}. We can see that when $\bar{\alpha}<-0.9$, even if the value of $\bar{b}$ is very small, there is still a resonance. As mentioned earlier, when $|\bar{\alpha}|$ is large, there are two wells for the effective potential, which can be seen from Fig.~\ref{figmodel2effectivep1}. It is because of the existence of these two potential wells that when $\bar{b}$ is small, the massive KK gravitons can be quasilocalized on the brane. However, when the value of $\bar{\alpha}$ increases from $-0.9$ to $0.00347$, a larger value of $\bar{b}$  is required for the existence of resonances. The emergence of these properties is due to the difference between $f(T)=T+\alpha T^{2}$ and GR.
\begin{figure}
	\centering
	{\label{model2alphab}
		\includegraphics[width=0.22\textwidth]{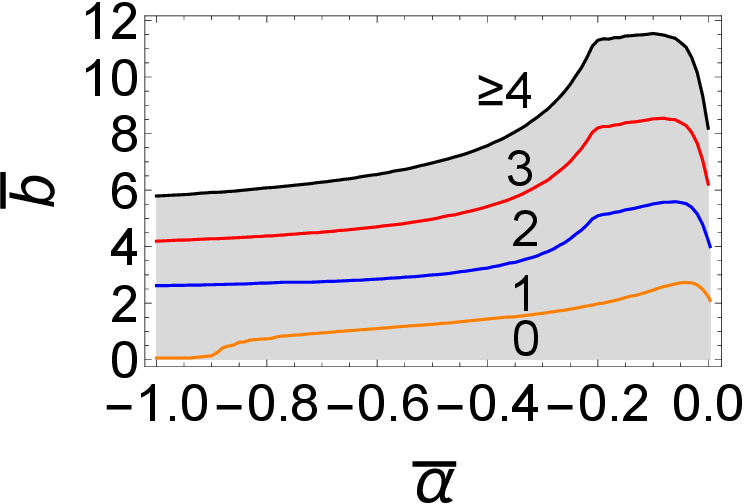}}
	\caption{The relation between the number of the resonances and the parameters $\bar{\alpha}$ and $\bar{b}$ for $f(T)=T+\frac{\bar{\alpha}}{k^{2}} T^{2}$. The number in each region represents the number of the resonances in this region.}\label{model2alphab}
\end{figure}

We know that there are positive correlations between the lifetime of the resonance and width and height of the effective potential. For the case of $f(T)=T$, we adjust the value of $\bar{b}$ to change the width of the potential well, but keep the height of the potential unchanged. For $f(T)=T+\alpha T^{2}$, the height of the potential and the lifetime of the first resonance increase with $|\bar{\alpha}|$.
When $|\bar{\alpha}|$ is large enough, the height of the potential increases very slowly, and when $\bar{\alpha}\rightarrow -\infty$, it approaches a finite value. In other words, the parameter $\bar{b}$ plays a key role in the lifetime of the first resonance.
This is similar to the case of $f(T)=T$. Thus, we also run into the dilemma as the GR case, i.e.,  the effective four-dimensional gravitational potential with such a large $\bar{b}$ may deviate from the squared inverse law at a large distance. So, let us consider another form of $f(T)$, which hopefully resolves this contradiction.

\subsection{Model 3: $f(T)=-T_{0}\left(e^{-\frac{T}{T_{0}}}-1\right)$}

\begin{figure}
	\centering
	\subfigure[~$\bar{T}_{0}=32$]{\label{figmodel3scalar1}
		\includegraphics[width=0.22\textwidth]{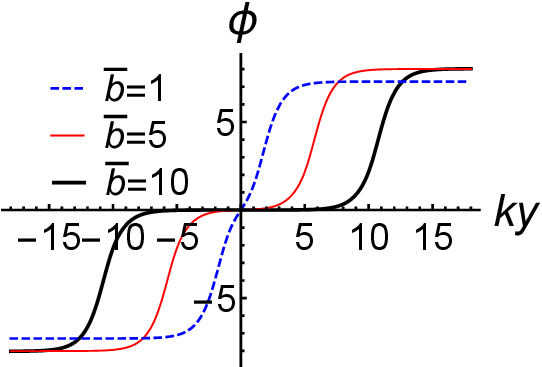}}
	\subfigure[~$\bar{b}=10$]{\label{figmodel3scalar2}
		\includegraphics[width=0.22\textwidth]{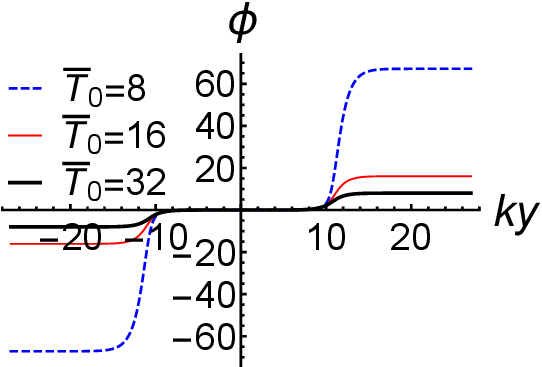}}
	\subfigure[~$\bar{T}_{0}=32$]{\label{figmodel3effectivep1}
		\includegraphics[width=0.22\textwidth]{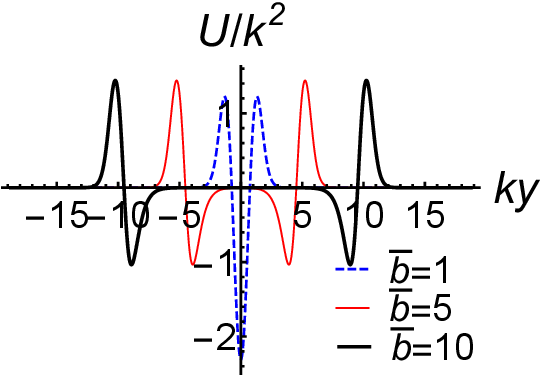}}
	\subfigure[~$\bar{b}=10$]{\label{figmodel3effectivep2}
		\includegraphics[width=0.22\textwidth]{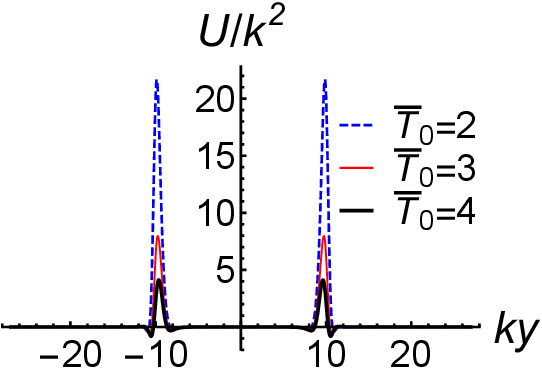}}
	\caption{Plots of the scalar field and the effective potential for $f(T)=-\frac{\bar{T}_{0}}{k^{2}}\left(e^{-\frac{k^{2}T}{\bar{T}_{0}}}-1\right)$.}\label{model3scalarandpotential}
\end{figure}

Lastly, we consider the model with
\begin{eqnarray}
	f(T)=-T_{0}\left(e^{-\frac{T}{T_{0}}}-1\right),\label{ft2}
\end{eqnarray}
where the mass dimension of the parameter $T_{0}$ is 2. For convenience, we define the dimensionless parameter $\bar{T}_{0}=T_{0}/k^{2}$. Then $f(T)$ can be rewritten as
\begin{eqnarray}
	f(T)=-\frac{\bar{T}_{0}}{k^{2}}\left(e^{-\frac{k^{2}T}{\bar{T}_{0}}}-1\right).
\end{eqnarray}
The expression (\ref{ft2}) can be expanded at $T=0$ as
\begin{eqnarray}
	f(T)=T-\frac{T^2}{2 T_0}+\frac{T^3}{6 T_0^{2}}+\cdots.\label{ft2expand}
\end{eqnarray}
It can be seen that the smaller $|T_0|$, the more deviation from GR. The warp factor is also considered as (\ref{warp factor}), and when $\bar{T}_{0}\rightarrow-\infty$ the model reverts back to general relativity.
In this model, Eq.~(\ref{phi field}) becomes
\begin{eqnarray}
	\phi'^{2}\!\!&=&\!\!\frac{3 k^2 \sinh ^4(2 b k)\text{sech}^8\big(k (y-b)\big) \text{sech}^8\big(k (b+y)\big)} {32 T_0 \Big(\tanh \big(k (y-b)\big)-\tanh \big(k (b+y)\big)\Big)^4} \nn\\
	&\times&\!\! \Bigg[2T_0 \Big(\cosh \big(2 k (y-b)\big)+\cosh \big(2 k (y+b)\big)+1\Big)\nn\\
	&-&\!\!96 k^2+T_0 \cosh (4 b k)+\left(96 k^2+T_0\right) \cosh (4 k y)\Bigg]\nn\\
	&\times&\!\! \Bigg[\cosh \big(2 k (y-b)\big)+\cosh \big(2 k (b+y)\big)+2\Bigg]\nn\\
	&\times&\!\! e^{\frac{12 k^2 }{T_0}\left[\tanh \left(k (y-b)\right)+\tanh \left(k (b+y)\right)\right]^2}.~~~~
	\label{model3scalarfield}
\end{eqnarray}
Same as before, from Eq.~(\ref{model3scalarfield}) we can find the numerical solution of the scalar field $\phi$. In order to ensure that the scalar field $\phi$ is real, the parameter $\bar{T}_{0}$ should satisfy $\bar{T}_{0}>0$ or $\bar{T}_{0}\leq-96$. To compare the deviation of these two branches of $T_{0}$ from GR, we rewrite expression (\ref{ft2expand}) as
\begin{eqnarray}
	f(T)=T\left(1-\frac{T}{2 T_0}+\frac{T^2}{6 T_0^{2}}+\cdots\right).
\end{eqnarray}
The $\frac{T}{2 T_0}$ term in the above equation is the dominant term deviating from GR.
From $T=-12 A'(y)^2$ and (\ref{warp factor}) we get that the range of $\frac{T}{k^2}$ is $-48\leq\frac{T}{k^2}\leq0$. For $\bar{T}_{0}\leq-96$, $0\leq \frac{T}{2T_{0}}\leq\frac{1}{4}$; for $\bar{T}_{0}>0$, $-\infty<\frac{T}{2T_{0}}\leq 0$. Obviously, the effect of torsion for the case $\bar{T}_{0}>0$ can be more significant than the case $\bar{T}_{0}\leq-96$, this is even more evident in the effective potential. Plots of the scalar field are shown in Figs.~\ref{figmodel3scalar1} and \ref{figmodel3scalar2}. It can be seen that the scalar field has the configuration of a double kink for a large $\bar{b}$. Similarly, we get the effective potential $U(z(y))$ through Eq.~(\ref{EffectivePotential}). Because the expression of the effective potential is complicated, we only show its plots. The width of the potential well increases with the parameter $\bar{b}$, and the height of the potential barrier decreases with the parameter $\bar{T}_{0}$, which can be seen from Figs.~\ref{figmodel3effectivep1} and \ref{figmodel3effectivep2}. The effective potential for $\bar{b}=10$, $\bar{T}_{0}=-96, -500$, and $\bar{T}_0\rightarrow-\infty$ (corresponding to GR) are shown in Fig.~\ref{model3potentialcompare}. We can see that all shapes of the effective potential for all values of $\bar{T}_0$ are almost the same. That is to say, when $\bar{T}_{0}\leq-96$, the torsion has an insignificant effect on the effective potential. Therefore, we only consider the case of $\bar{T}_{0}>0$.
\begin{figure}
	\centering
	\subfigure[~$\bar{b}=10$]{\label{model3potentialcompare}
		\includegraphics[width=0.22\textwidth]{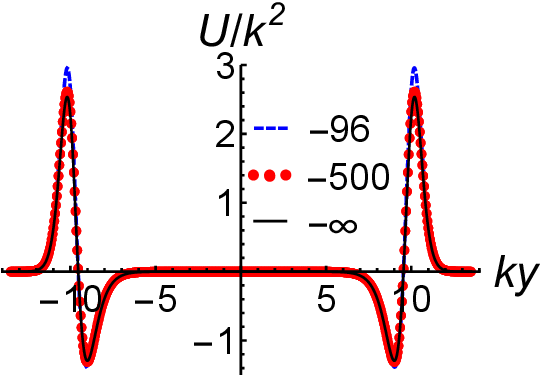}}
	\caption{Plot of the effective potential for $f(T)=-\frac{\bar{T}_{0}}{k^{2}}\left(e^{-\frac{k^{2}T}{\bar{T}_{0}}}-1\right)$. The parameter $\bar{T}_{0}$ is set to $\bar{T}_{0}=-96, -500$, and $\bar{T}_{0}\rightarrow-\infty$.}\label{model3potentialcompare}
\end{figure}

\begin{figure}
	\centering
	\subfigure[~$\bar{T}_{0}=2$]{\label{figmodel3p3}
		\includegraphics[width=0.22\textwidth]{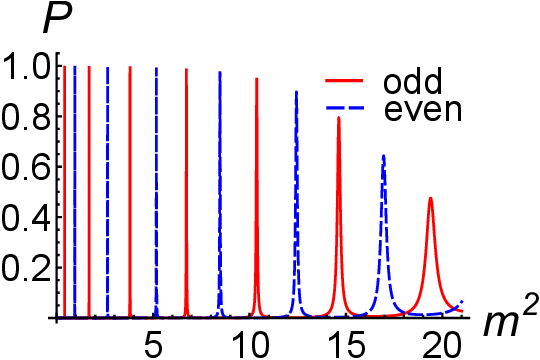}}
	\subfigure[~$\bar{T}_{0}=3$]{\label{figmodel3p2}
		\includegraphics[width=0.22\textwidth]{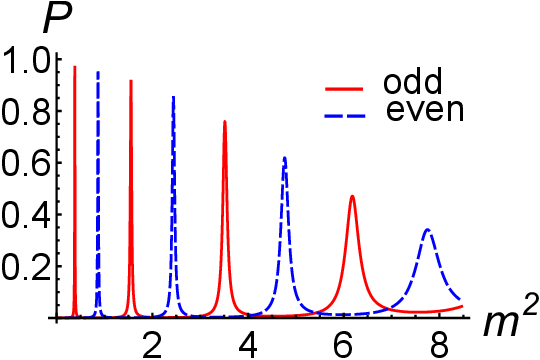}}
	\subfigure[~$\bar{T}_{0}=4$]{\label{figmodel3p1}
		\includegraphics[width=0.22\textwidth]{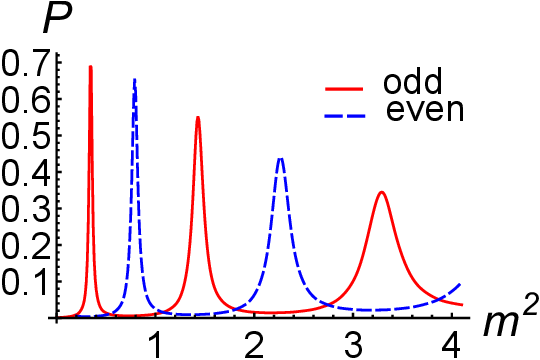}}
	\caption{The relative probability $P$ for different values of $\bar{T}_{0}$ for $f(T)=-\frac{\bar{T}_{0}}{k^{2}}\left(e^{-\frac{k^{2}T}{\bar{T}_{0}}}-1\right)$. The red and dashed blue lines correspond to the odd and even parities, respectively. The parameter is set to $\bar{b}=10$.}\label{model3P}
\end{figure}

The relative probability $P(m^{2})$ of the gravitational  resonances for $\bar{b}=10$, $\bar{T}_{0}=2, 3$, and $4$ are shown in Fig.~\ref{model3P}. We find that the number of the resonances decreases with $\bar{T}_{0}$.
The relation between the number of the resonances and the parameters $\bar{T}_{0}$ and $\bar{b}$ is shown in Fig.~\ref{model3Tb}. We can see that, when $\bar{T}_{0}<6$, even if the value of $\bar{b}$ is very small, there is still a resonance. This is similar to the case of $f(T)=T+\alpha T^{2}$.
\begin{figure}
	\centering
	{\label{model3Tb}
		\includegraphics[width=0.22\textwidth]{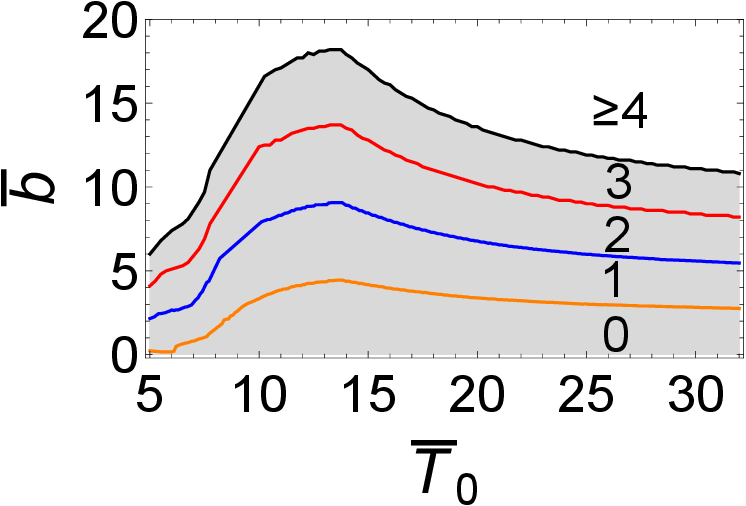}}
	\caption{The relation between the number of the resonances and the parameters $\bar{T}_{0}$ and $\bar{b}$ for $f(T)=-\frac{\bar{T}_{0}}{k^{2}}\left(e^{-\frac{k^{2}T}{\bar{T}_{0}}}-1\right)$.}\label{model3Tb}
\end{figure}

\begin{figure}
	\centering
	\subfigure[~$\bar{m}_{1}(\bar{T}_{0})$]{\label{2tm}
		\includegraphics[width=0.22\textwidth]{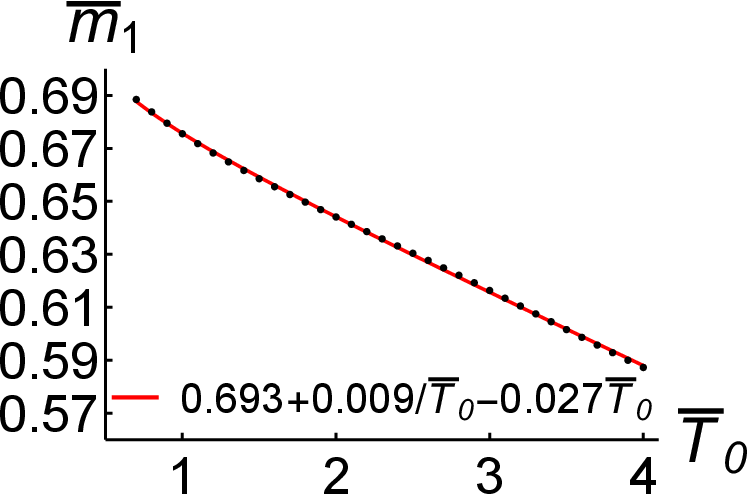}}
	\subfigure[~$\bar{\tau}_{1}(\bar{T}_{0})$]{\label{2tlife}
		\includegraphics[width=0.22\textwidth]{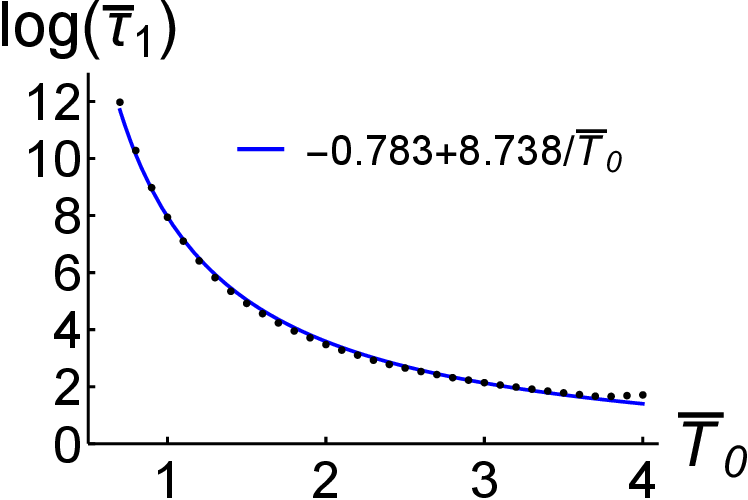}}
	\caption{The relations of the scaled mass $\bar{m}_{1}$ and the scaled lifetime $\bar{\tau}_{1}$ of the first resonance with the parameter $\bar{T}_{0}$ for $f(T)=-\frac{\bar{T}_{0}}{k^{2}}\left(e^{-\frac{k^{2}T}{\bar{T}_{0}}}-1\right)$. The dots are calculated values while the red and blue lines are fit functions. The parameter is set to $\bar{b}=10$.}\label{model3mt}
\end{figure}

The relations of the scaled mass $\bar{m}_{1}$ and the scaled lifetime $\bar{\tau}_{1}$ of the first resonance with the parameter $\bar{T}_{0}$ are shown in Fig.~\ref{model3mt}. It can be seen that, both the scaled mass $\bar{m}_{1}$ and the scaled lifetime $\bar{\tau}_{1}$ of the first resonance decrease with $\bar{T}_{0}$. Similarly, the KK graviton resonances with long enough lifetime could be considered as one of the candidates for dark matter.
The scaled mass $\bar{m}_{1}$ and the scaled lifetime $\bar{\tau}_{1}$ can be fitted as two functions of the parameter $\bar{T}_{0}$, and the fit functions can be expressed as
\begin{eqnarray}
	\bar{m}_{1}&=&0.693+\frac{0.009}{\bar{T}_{0}}-0.027\bar{T}_{0}, \label{fitm2}\\
	\log(\bar{\tau}_{1})&=&-0.783+\frac{8.738}{\bar{T}_{0}}\label{fitt2}.
\end{eqnarray}

\begin{figure}
	\centering
	\subfigure[~$k-\bar{T}_{0}$]{\label{figmodel3limitkT}
		\includegraphics[width=0.23\textwidth]{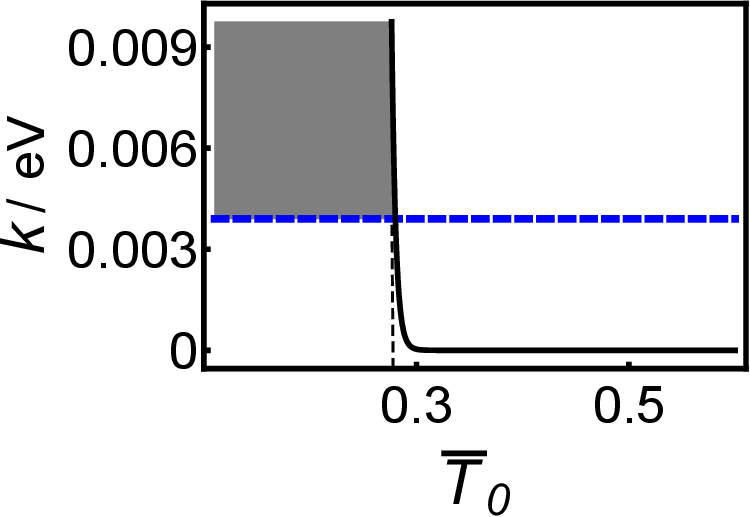}}
	\subfigure[~$m_{1}-\bar{T}_{0}$]{\label{figmodel3limitmT}
		\includegraphics[width=0.23\textwidth]{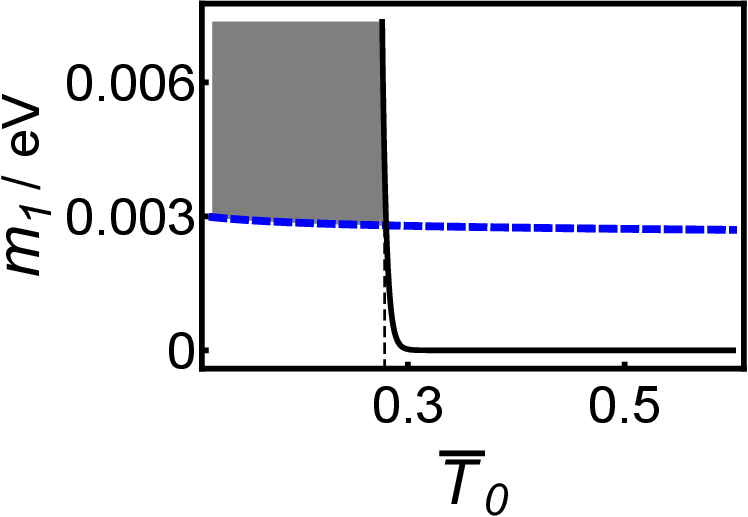}}
	\caption{The limit range of the parameter $k$ and the corresponding range of the first resonance mass $m_{1}$. The shadow region of the left panel is the limit range of the parameter $k$ and the shadow region of the right panel is the corresponding range of the first resonance mass $m_{1}$. The black solid lines are the limits that the lifetime of the first resonance should be longer than the age of the universe and the blue dashed lines are limits from $1/k$ should be less than $50~{\rm\mu m}$.}\label{model3limit}
\end{figure}

On the other hand, since we have no analytical formulation for the integral expression (\ref{conditionMall}) for $f(T)=-T_{0}\left(e^{-\frac{T}{T_{0}}}-1\right)$, we cannot obtain an analytical constraint for the five-dimensional fundamental scale $M_5$. In this case, we consider the constraint imposed by the test of the gravitational inverse-square law. According to the method of Ref.~\cite{Csaki:2000fc}, for the brane world model considered here, the four-dimensional gravitational potential is
\begin{eqnarray}
	\label{correction potential}
	V(r) \sim G_N \frac{M}{r}
	\left[1+ \frac{C}{(kr)^2} \right],
\end{eqnarray}
where $C$ is a dimensionless constant determined by the structure of the brane. The correction term is $G_{N}\frac{CM}{k^2r^3}$. For the case of $f(T)=-T_{0}\left(e^{-\frac{T}{T_{0}}}-1\right)$, the correction occurs at the scale of $r\sim 1/k$.  Experimentally, in the recent experiments of the test of the gravitational inverse-square law, the usual Newtonian potential still holds down to the scale of about $50~{\rm\mu m}$~\cite{Tan:2016vwu,Lee:2020zjt,Tan:2020vpf}. Therefore, $1/k$ should be less than $50~{\rm\mu m}$, this is equivalent to
\begin{eqnarray}
	k > 4~\times{10^{-3}}~\text{eV}. \label{correction potential limitk}
\end{eqnarray}
On the other hand, the restriction of the lifetime of the first resonance on the parameter $k$ is given by
\begin{equation}
	k \lesssim 1.5 \times{10^{\frac{8.738}{\bar{T}_{0}}-33.783}}~\text{eV}.\label{limitkt}
\end{equation}
By combining the fit function \eqref{fitm2} and the two conditions \eqref{correction potential limitk}, \eqref{limitkt}, the restricted expressions of the mass of the first resonance $m_{1}$ with the parameter $\bar{T}_{0}$ can be obtained
\begin{eqnarray}
	m_1&>&\left(2.8-0.1\bar{T}_{0}+\frac{0.04}{\bar{T}_{0}}\right)\times10^{-3}~\text{eV}\label{conditionm1},\\
	m_1&\lesssim&\left(1.0-0.04\bar{T}_{0}+\frac{0.014}{\bar{T}_{0}}\right)\times10^{\frac{8.738}{\bar{T}_{0}}-33.783} \text{eV}\label{conditionm2}.\nn\\
\end{eqnarray}
The available ranges of the parameters $k$ and $m_{1}$ are shown in Fig.~\ref{model3limit}. From Fig.~\ref{figmodel3limitkT}, we can see that only if $\bar{T}_{0}<0.28$, the two restricted conditions \eqref{correction potential limitk} and \eqref{limitkt} of $k$ could be satisfied, which means that the parameter $\bar{T}_{0}$ has an upper bound. From Fig.~\ref{figmodel3limitmT}, we can see that the first resonance mass $m_{1}$ has a lower bound, i.e., $m_{1}\gtrsim 2.8\times 10^{-3}\text{eV}$. It can be found that for model 3, the first resonance as a candidate for dark matter does not have the previous confliction of model 1 and model 2. The main reason is that the key role that determines the lifetime of the first resonance is no longer $\bar{b}$ but $\bar{T_{0}}$. This reminds us that, if the effect of the torsion is large enough, other forms of $f(T)$ can also produce a long-lived resonance that satisfies the conditions to be a candidate of dark matter.

\section{Conclusion and discussion}
\label{Conclusion}

In this work, we investigated the gravitational resonances in various $f(T)$-brane models. The branes are generated by a canonical scalar field. We first reviewed the tensor perturbations of $f(T)$-branes. Then we discussed the gravitational resonances in three $f(T)$-brane models with the same warp factor $(\ref{warp factor})$.

In model 1, we considered $f(T)=T$ and obtained the thick brane solution, which is equivalent to GR. We found that with the increase of the parameter $\bar{b}$, the double kink structure appears in the scalar field, which generally corresponds to the appearance of sub branes and resonances. The influence of the parameter $\bar{b}$ on the gravitational resonances was analyzed. The result shows that, only when $\bar{b}>1$, the gravitational resonances could exist. The mass $m_1$ of the first resonance decreases with the parameter $\bar{b}$. The lifetime $\tau_1$ of the first resonance and the number of resonances increase with the parameter $\bar{b}$. If $\bar{b}$ is large enough, the lifetime of the first resonance will be long enough as the age of our universe. This indicates that the first gravitational resonance could be one of the candidates for dark matter, which would lead to some constraints on the parameters $k$ and $\bar{b}$, i.e., $k\gtrsim9.5\times{10^{-7}}~\text{eV}$, $\bar{b}>7.9~\times10^{9}$. And the corresponding available range of the first resonance mass is $m_{1}\gtrsim 4\times 10^{-16}~\text{eV}$. But there is a problem, with such a large $\bar{b}$, the effective four-dimensional gravitational potential may deviate from the squared inverse law at a large distance.

In model 2, we considered $f(T)=T+\alpha T^{2}$. When the parameter $\bar{\alpha}$ changes, the peak of the effective potential changes from two to four, suggesting that $f(T)$-brane has more abundant internal structure and properties than GR-brane. The parameter $\bar{\alpha}$ can affect the resonance spectrum and the lifetime of resonances. The influence of the parameter $\bar{b}$ is the same as $f(T)=T$. But the parameter $\bar{b}$ plays a key role in the lifetime of the first resonance. The relation between the number of the resonances and the parameters $\bar{\alpha}$ and $\bar{b}$ was obtained, which can be seen from Fig.~\ref{model2alphab}. This can reflect the characteristics of the extra dimension. We found that for large $|\bar{\alpha}|$, there is a gravitational resonance even for small $\bar{b}$, which is different from GR.  Unfortunately, when considering the resonance as a candidate for dark matter, this model has the same problem as GR.

In model 3, $f(T)=-T_{0}\left(e^{-\frac{T}{T_{0}}}-1\right)$ was considered. We analyzed the effects of the parameters $\bar{T}_{0}$ and $\bar{b}$ on the effective potential, and found that the parameter $\bar{T}_{0}$ significantly changes the height of the potential barrier. Then, we got the relation between the number of the resonances and the parameters $\bar{T}_{0}$ and $\bar{b}$ (see Fig.~\ref{model3Tb}). The influence of the parameter $\bar{T}_{0}$ on the gravitational resonances was analyzed by taking $\bar{b}=10$. The result shows that the lifetime $\tau_1$ of the first resonance decreases with the parameter $\bar{T}_{0}$. If the lifetime of the first resonance exceeds the age of our universe, the parameters $k$, $\bar{T}_{0}$, and mass of the first resonance must satisfy $k\gtrsim4\times{10^{-3}}~\text{eV}$, $\bar{T}_{0}<0.28$, and $m_{1}\gtrsim 2.8\times 10^{-3}~\text{eV}$. We can see that model 3 does not have the two problems in model 1 and model 2. This indicates that the first resonance could be a candidate for dark matter as $\bar{T}_{0}<0.28$, for which the $f(T)$ theory is very different from GR. Note that, if we choose a larger value of $\bar{b}$, the constraint on $\bar{T}_{0}$ will be further relaxed. In a word, if we want the first gravitational resonance to be a candidate of dark matter, the effect of torsion should be significant. That is, other forms of $f(T)$ gravity may also satisfy the conditions of first resonance as a candidate for dark matter, which is worth further study.

In this paper, we only considered the KK gravitons in $f(T)$ gravity as the candidate for dark matter and investigated the corresponding properties of such resonances. In fact, KK fermions and KK vector particles, etc., may also be candidates for dark matter, which deserves further study.
\section*{Acknowledgements}
We are thankful to T.T.~Sui and Y.M.~Xu for useful discussions. This work was supported by the National Key Research and Development Program of China (Grant No. 2020YFC2201400), the National Natural Science Foundation of China (Grants No.~11875151 and No.~11522541), the 111 Project (Grant No. B20063), and the Fundamental Research Funds for the Central Universities (Grants No. lzujbky-2019-it21 and No.~lzujbky-2019-ct06).


\begin{thebibliography}{99}
	
	\bibitem{kaluza1921unitatsproblem}
	T.~Kaluza, \emph{Zum unit{\"a}tsproblem der physik}, {\emph{Sitzungsber.
			Preuss. Akad. Wiss. Berlin (Math. Phys.)} {\bfseries 27} (1921) 966}.
	
	\bibitem{Klein:1926tv}
	O.~Klein, \emph{{Quantum Theory and Five-Dimensional Theory of Relativity. (In
			German and English)}}, {\emph{Z.
			Phys.} {\bfseries 37} (1926) 895}.
	
	\bibitem{Akama:1982jy}
	K.~Akama, \emph{{An Early Proposal of `Brane World'}}, {\emph{Lect. Notes
			Phys.} {\bfseries 176} (1982) 267},
	[{{\ttfamily arXiv:hep-th/0001113}}].
	
	\bibitem{Rubakov:1983bb}
	V.~A. Rubakov and M.~E. Shaposhnikov, \emph{{Do We Live Inside a Domain
			Wall?}} {\emph{Phys.
			Lett. B} {\bfseries 125} (1983) 136}.
	
	\bibitem{Rubakov:1983bz}
	V.~A. Rubakov and M.~E. Shaposhnikov, \emph{{Extra space-time Dimensions:
			Towards a Solution to the Cosmological Constant Problem}},
	{\emph{Phys. Lett. B}
		{\bfseries 125} (1983) 139}.
	
	\bibitem{ArkaniHamed:1998rs}
	N.~Arkani-Hamed, S.~Dimopoulos, and G.~R. Dvali, \emph{{The Hierarchy problem
			and new dimensions at a millimeter}},
	{\emph{Phys. Lett. B}
		{\bfseries 429} (1998) 263},
	[{{\ttfamily arXiv:hep-ph/9803315}}].
	
	\bibitem{Randall:1999ee}
	L.~Randall and R.~Sundrum, \emph{{A Large mass hierarchy from a small extra dimension}},
	{\emph{Phys. Rev. Lett.}
		{\bfseries 83} (1999) 3370},
	[{{\ttfamily arXiv:hep-ph/9905221}}].
	
	\bibitem{Randall:1999vf}
	L.~Randall and R.~Sundrum, \emph{{An Alternative to compactification}},
	{\emph{Phys. Rev. Lett.}
		{\bfseries 83} (1999) 4690},
	[{{\ttfamily arXiv:hep-th/9906064}}].
	
	\bibitem{Goldberger:1999uk}
	W.~D. Goldberger and M.~B. Wise, \emph{{Modulus stabilization with bulk
			fields}}, {\emph{Phys.
			Rev. Lett.} {\bfseries 83} (1999) 4922},
	[{{\ttfamily arXiv:hep-ph/9907447}}].
	
	\bibitem{Gremm:1999pj}
	M.~Gremm, \emph{{Four-dimensional gravity on a thick domain wall}},
	{\emph{Phys. Lett. B}
		{\bfseries 478} (2000) 434},
	[{{\ttfamily arXiv:hep-th/9912060}}].
	
	\bibitem{DeWolfe:1999cp}
	O.~DeWolfe, D.~Z. Freedman, S.~S. Gubser, and A.~Karch, \emph{{Modeling the
			fifth-dimension with scalars and gravity}},
	{\emph{Phys. Rev. D}
		{\bfseries 62} (2000) 046008},
	[{{\ttfamily arXiv:hep-th/9909134}}].
	
	\bibitem{Bazeia:2008zx}
	D.~Bazeia, A.~R. Gomes, L.~Losano, and R.~Menezes, \emph{{Braneworld Models of
			Scalar Fields with Generalized Dynamics}},
	{\emph{Phys. Lett. B}
		{\bfseries 671} (2009) 402},
	[{{\ttfamily arXiv:0808.1815}}].
	
	
	\bibitem{Charmousis:2001hg}
	C.~Charmousis, R.~Emparan, and R.~Gregory, \emph{{Selfgravity of brane worlds: A
			New hierarchy twist}},
	{\emph{JHEP} {\bfseries
			05} (2001) 026}, [{{\ttfamily
			arXiv:hep-th/0101198}}].
	
	\bibitem{Arias:2002ew}
	O.~Arias, R.~Cardenas, and I.~Quiros, \emph{{Thick brane worlds arising from
			pure geometry}},
	{\emph{Nucl. Phys. B}
		{\bfseries 643} (2002) 187},
	[{{\ttfamily arXiv:hep-th/0202130}}].
	
	\bibitem{Barcelo:2003wq}
	C.~Barcelo, C.~Germani, and C.~F. Sopuerta, \emph{{On the thin shell limit of
			branes in the presence of Gauss-Bonnet interactions}},
	{\emph{Phys. Rev. D}
		{\bfseries 68} (2003) 104007},
	[{{\ttfamily arXiv:gr-qc/0306072}}].
	
	\bibitem{Bazeia:2004dh}
	D.~Bazeia and A.~R.~Gomes, \emph{{Bloch brane}}, {\emph{JHEP} {\bfseries 05} (2004) 012}, [{{\ttfamily arXiv:hep-th/0403141}}].
	
	\bibitem{CastilloFelisola:2004eg}
	O.~Castillo-Felisola, A.~Melfo, N.~Pantoja, and A.~Ramirez, \emph{{Localizing gravity on exotic thick three-branes}}, {\emph{Phys. Rev. D}
		{\bfseries 70} (2004) 104029},
	[{{\ttfamily arXiv:hep-th/0404083}}].
	
	\bibitem{BarbosaCendejas:2005kn}
	N.~Barbosa-Cendejas and A.~Herrera-Aguilar, \emph{{4D gravity localized in non
			$Z_2$ symmetric thick branes}},
	{\emph{JHEP} {\bfseries
			10} (2005) 101}, [{{\ttfamily
			arXiv:hep-th/0511050}}].
	
	\bibitem{Koerber:2008rx}
	P.~Koerber, D.~Lust, and D.~Tsimpis, \emph{{Type IIA AdS$_4$ compactifications on
			cosets, interpolations and domain walls}},
	{\emph{JHEP} {\bfseries
			07} (2008) 017}, [{{\ttfamily
			arXiv:0804.0614}}].
	
	\bibitem{BarbosaCendejas:2007vp}
	N.~Barbosa-Cendejas, A.~Herrera-Aguilar, M.~A. Reyes~Santos, and C.~Schubert,
	\emph{{Mass gap for gravity localized on Weyl thick branes}},
	{\emph{Phys. Rev. D}
		{\bfseries 77} (2008) 126013},
	[{{\ttfamily arXiv:0709.3552}}].
	
	\bibitem{Johnson:2008kc}
	M.~C. Johnson and M.~Larfors, \emph{{Field dynamics and tunneling in a flux
			landscape}}, {\emph{Phys.
			Rev. D} {\bfseries 78} (2008) 083534},
	[{{\ttfamily arXiv:0805.3705}}].
	
	
	
	\bibitem{Liu:2011wi}
	Y.-X. Liu, Y.~Zhong, Z.-H. Zhao, and H.-T. Li, \emph{{Domain wall brane in
			squared curvature gravity}},
	{\emph{JHEP} {\bfseries 06}
		(2011) 135}, [{{\ttfamily
			arXiv:1104.3188}}].
	
	
	\bibitem{Kanno:2004nr}
	S.~Kanno and J.~Soda, \emph{{Quasi-thick codimension 2 braneworld}},
	{\emph{JCAP} {\bfseries
			0407} (2004) 002}, [{{\ttfamily
			arXiv:hep-th/0404207}}].
	
	\bibitem{Chumbes:2011zt}
	A.~E.~R. Chumbes, J.~M. Hoff~da Silva, and M.~B. Hott, \emph{{A model to
			localize gauge and tensor fields on thick branes}},
	{\emph{Phys. Rev. D}
		{\bfseries 85} (2012) 085003},
	[{{\ttfamily arXiv:1108.3821}}].
	
	\bibitem{Andrianov:2012ae}
	A.~A. Andrianov, V.~A. Andrianov, and O.~O. Novikov, \emph{{Localization of
			scalar fields on self-gravitating thick branes}},
	{\emph{Phys. Part. Nucl.}
		{\bfseries 44} (2013) 190},
	[{{\ttfamily arXiv:1210.3698}}].
	
	\bibitem{Kulaxizi:2014yxa}
	M.~Kulaxizi and R.~Rahman, \emph{{Higher-Spin Modes in a Domain-Wall
			Universe}}, {\emph{JHEP}
		{\bfseries 10} (2014) 193},
	[{{\ttfamily arXiv:1409.1942}}].
	
	\bibitem{Dutra:2014xla}
	A.~de~Souza~Dutra, G.~P. de~Brito, and J.~M. Hoff~da Silva, \emph{{Method for
			obtaining thick brane models}},
	{\emph{Phys. Rev. D}
		{\bfseries 91} (2015) 086016},
	[{{\ttfamily arXiv:1412.5543}}].
	
	\bibitem{Almeida:2009jc}
	C.~A.~S. Almeida, M.~M. Ferreira, A.~R. Gomes, and R.~Casana,
	\emph{{Fermion localization and resonances on two-field thick branes}},
	{\emph{Phys. Rev. D}
		{\bfseries 79} (2009) 125022},
	[{{\ttfamily arXiv:0901.3543}}].
	
	
	\bibitem{Karam:2018squ}
	A.~Karam, A.~Lykkas, and K.~Tamvakis,
	\emph{{Frame-invariant approach to higher-dimensional scalar-tensor gravity}},
	{\emph{Phys. Rev. D} {\bfseries 97} (2018) 124036},
	[{{\ttfamily arXiv:1803.04960}}].
	
	\bibitem{Dzhunushaliev:2009va}
	V.~Dzhunushaliev, V.~Folomeev, and M.~Minamitsuji, \emph{{Thick brane solutions}},
	{\emph{Rept. Prog.
			Phys.} {\bfseries 73} (2010) 066901},
	[{{\ttfamily arXiv:0904.1775}}].
	
	\bibitem{Zhou:2017xaq}
	X.-N~Zhou, Y.-Z~Du, H.~Yu, and Y.-X~Liu, \emph{{Localization of Gravitino Field on $f(R)$ Thick Branes}},
	{\emph{Sci. China Phys. Mech. Astron.} {\bfseries 61} (2018) 110411},
	[{{\ttfamily arXiv:1703.10805}}].
	
	\bibitem{Zhong:2014kha}
	Y.~Zhong and Y.-X. Liu, \emph{{$K$-field kinks: stability, exact solutions and
			new features}}, {\emph{JHEP}
		{\bfseries 10} (2014) 041},
	[{{\ttfamily arXiv:1408.4511}}].
	
	\bibitem{Navarro:2004di}
	I.~Navarro and J.~Santiago, \emph{{Gravity on codimension 2 brane worlds}},
	{\emph{JHEP} {\bfseries
			02} (2005) 007}, [{{\ttfamily
			arXiv:hep-th/0411250}}].
	
	\bibitem{deSouzaDutra:2008gm}
	A.~de~Souza~Dutra, A.~C.~Amaro de~Faria, and M.~Hott, \emph{{Degenerate and
			critical Bloch branes}},
	{\emph{Phys. Rev. D}
		{\bfseries 78} (2008) 043526},
	[{{\ttfamily arXiv:0807.0586}}].
	
	\bibitem{Xie:2019jkq}
	Q.-Y. Xie, Z.-H. Zhao, J.~Yang, and K.~Yang, \emph{{Fermion Localization and
			Degenerate Resonances on Brane Array}},
	{\emph{Class. Quant. Grav.}
		{\bfseries 37} (2020) 025012},
	[{{\ttfamily arXiv:1901.11253}}].
	
	\bibitem{Bajc:1999mh}
	B.~Bajc and G.~Gabadadze, \emph{{Localization of matter and cosmological
			constant on a brane in anti-de Sitter space}},
	{\emph{Phys. Lett. B}
		{\bfseries 474} (2000) 282},
	[{{\ttfamily arXiv:hep-th/9912232}}].
	
	\bibitem{Bagger:2004rr}
	J.~A. Bagger and D.~V. Belyaev, \emph{{Brane-localized Goldstone fermions in
			bulk supergravity}},
	{\emph{Phys. Rev. D}
		{\bfseries 72} (2005) 065007},
	[{{\ttfamily arXiv:hep-th/0406126}}].
	
	\bibitem{Ringeval:2001cq}
	C.~Ringeval, P.~Peter, and J.-P. Uzan, \emph{{Localization of massive fermions
			on the brane}},
	{\emph{Phys. Rev. D}
		{\bfseries 65} (2002) 044016},
	[{{\ttfamily arXiv:hep-th/0109194}}].
	
	\bibitem{Liu:2007ku}
	Y.-X. Liu, X.-H. Zhang, L.-D. Zhang, and Y.-S. Duan, \emph{{Localization of
			Matters on Pure Geometrical Thick Branes}},
	{\emph{JHEP} {\bfseries
			02} (2008) 067}, [{{\ttfamily
			arXiv:0708.0065}}].
	
	\bibitem{Liu:2008wd}
	Y.-X. Liu, L.-D. Zhang, S.-W. Wei, and Y.-S. Duan, \emph{{Localization and Mass
			Spectrum of Matters on Weyl Thick Branes}},
	{\emph{JHEP} {\bfseries
			08} (2008) 041}, [{{\ttfamily
			arXiv:0803.0098}}].
	
	\bibitem{Ghoroku:2001zu}
	K.~Ghoroku and A.~Nakamura, \emph{{Massive vector trapping as a gauge boson on
			a brane}}, {\emph{Phys.
			Rev. D} {\bfseries 65} (2002) 084017},
	[{{\ttfamily arXiv:hep-th/0106145}}].
	
	
	\bibitem{Csaki:2000fc}
	C.~Csaki, J.~Erlich, T.~J. Hollowood, and Y.~Shirman, \emph{{Universal aspects
			of gravity localized on thick branes}},
	{\emph{Nucl. Phys. B}
		{\bfseries 581} (2000) 309},
	[{{\ttfamily arXiv:hep-th/0001033}}].
	
	\bibitem{Liu:2009ve}
	Y.-X. Liu, J.~Yang, Z.-H. Zhao, C.-E. Fu, and Y.-S. Duan,
	\emph{{Fermion Localization and Resonances on A de Sitter Thick Brane}},
	{\emph{Phys. Rev. D}{\bfseries 80} (2009) 065019},
	[{{\ttfamily arXiv:0904.1785}}].
	
	\bibitem{Cooper:1994eh}
	F.~Cooper, A.~Khare, and U.~Sukhatme,
	\emph{{Supersymmetry and quantum mechanics}},
	{\emph{Phys. Rept. }{\bfseries 251} (1995) 267},
	[{{\ttfamily arXiv:hep-th/9405029}}].
	
	\bibitem{Cruz:2013uwa}
	W.~T. Cruz, L.~J.~S. Sousa, R.~V. Maluf, and C.~A.~S. Almeida, \emph{{Graviton
			resonances on two-field thick branes}},
	{\emph{Phys. Lett. B}
		{\bfseries 730} (2014) 314},
	[{{\ttfamily arXiv:1310.4085}}].
	
	\bibitem{Xu:2014jda}
	Z.-G. Xu, Y.~Zhong, H.~Yu, and Y.-X. Liu, \emph{{The structure of $f(R)$-brane
			model}}, {\emph{Eur.
			Phys. J. C} {\bfseries 75} (2015) 368},
	[{{\ttfamily arXiv:1405.6277}}].
	
	\bibitem{Csaki:2000pp}
	C.~Csaki, J.~Erlich, and T.~J. Hollowood, \emph{{Quasilocalization of gravity by
			resonant modes}},
	{\emph{Phys. Rev. Lett.}
		{\bfseries 84} (2000) 5932},
	[{{\ttfamily arXiv:hep-th/0002161}}].
	
	\bibitem{Gregory:2000jc}
	R.~Gregory, V.~A. Rubakov, and S.~M. Sibiryakov, \emph{{Opening up extra
			dimensions at ultra large scales}},
	{\emph{Phys. Rev. Lett.}
		{\bfseries 84} (2000) 5928},
	[{{\ttfamily arXiv:hep-th/0002072}}].
	
	
	\bibitem{Zhang:2016ksq}
	Y.-P.~Zhang, Y.-Z.~Du, W.-D.~Guo and Y.-X.~Liu,
	\emph{{Resonance spectrum of a bulk fermion on branes}},
	{\emph{Phys. Rev. D}  {\bfseries93} (2016) 065042},
	[{{\ttfamily arXiv:1601.05852}}].
	
	
	\bibitem{Sui:2020fty}
	T.-T. Sui, W.-D. Guo, Q.-Y. Xie and Y.-X. Liu,
	\emph{{Generalized geometrical coupling for vector field localization on thick brane in asymptotic Anti-de
			Sitter spacetime}},
	{\emph{Phys. Rev. D}  {\bfseries101} (2020) 055031},
	[{{\ttfamily arXiv:2001.02154}}].
	
	\bibitem{Nollert:1999ji}
	H.-P. Nollert, \emph{{TOPICAL REVIEW: Quasinormal modes: the characteristic
			`sound' of black holes and neutron stars}},
	{\emph{Class. Quant.
			Grav.} {\bfseries 16} (1999) R159}.
	
	\bibitem{Hayashi:1979qx}
	K.~Hayashi and T.~Shirafuji, \emph{{New General Relativity}},
	{\emph{Phys. Rev. D}
		{\bfseries 19} (1979) 3524}.
	
	\bibitem{Sousa:2007zc}
	A.~A. Sousa, J.~S. Moura, and R.~B. Pereira, \emph{{Energy in an Expanding
			Universe in the Teleparallel Geometry}},
	{\emph{Braz. J.
			Phys.} {\bfseries 40} (2010) 1},
	[{{\ttfamily arXiv:gr-qc/0702109}}].
	
	\bibitem{Bengochea:2008gz}
	G.~R. Bengochea and R.~Ferraro, \emph{{Dark torsion as the cosmic speed-up}},
	{\emph{Phys. Rev. D}
		{\bfseries 79} (2009) 124019},
	[{{\ttfamily arXiv:0812.1205}}].
	
	
	\bibitem{Chen:2010va}
	S.-H. Chen, J.-B. Dent, S. Dutta, and E.~N. Saridakis,
	\emph{{Cosmological perturbations in f(T) gravity}},
	{\emph{Phys. Rev. D}
		{\bfseries 83} (2011) 023508},
	[{{\ttfamily arXiv:1008.1250}}].
	
	
	\bibitem{Farrugia:2018gyz}
	G.~Farrugia, J.~Levi Said, V.~Gakis, and E.~N.~Saridakis,
	\emph{{Gravitational Waves in Modified Teleparallel Theories}},
	{\emph{Phys. Rev. D}
		{\bfseries 97} (2018)  124064},
	[{{\ttfamily arXiv:1804.07365}}].
	
	
	\bibitem{Cai:2015emx}
	Y.-F. Cai, S.~Capozziello, M.~De~Laurentis, and E.~N. Saridakis, \emph{{$f(T)$
			teleparallel gravity and cosmology}},
	{\emph{Rept. Prog.
			Phys.} {\bfseries 79} (2016) 106901},
	[{{\ttfamily
			arXiv:1511.07586}}].
	
	\bibitem{Ferraro:2006jd}
	R.~Ferraro and F.~Fiorini, \emph{{Modified teleparallel gravity: Inflation
			without inflaton}},
	{\emph{Phys. Rev. D}
		{\bfseries 75} (2007) 084031},
	[{{\ttfamily arXiv:gr-qc/0610067}}].
	
	\bibitem{Bamba:2012vg}
	K.~Bamba, R.~Myrzakulov, S.~Nojiri, and S.~D. Odintsov, \emph{{Reconstruction of
			$f(T)$ gravity: Rip cosmology, finite-time future singularities and
			thermodynamics}},
	{\emph{Phys. Rev. D}
		{\bfseries 85} (2012) 104036},
	[{{\ttfamily arXiv:1202.4057}}].
	
	\bibitem{Fiorini:2013hva}
	F.~Fiorini, P.~A. Gonzalez, and Y.~Vasquez, \emph{{Compact extra dimensions in
			cosmologies with f(T) structure}},
	{\emph{Phys. Rev. D}
		{\bfseries 89} (2014) 024028},
	[{{\ttfamily arXiv:1304.1912}}].
	
	\bibitem{Geng:2014nfa}
	C.-Q. Geng, C.~Lai, L.-W. Luo, and H.-H. Tseng, \emph{{Kaluza Klein theory for
			teleparallel gravity}},
	{\emph{Phys. Lett. B}
		{\bfseries 737} (2014) 248},
	[{{\ttfamily arXiv:1409.1018}}].
	
	
	\bibitem{Li:2018ixg}
	C.~Li, Y.~Cai, Y.-F.~Cai, and E.~N.~Saridakis,
	\emph{{The effective field theory approach of teleparallel gravity, $f(T)$ gravity and beyond}},
	{\emph{JCAP}
		{\bfseries 10} (2018) 001},
	[{{\ttfamily arXiv:1803.09818}}].
	
	
	\bibitem{Yang:2012hu}
	J.~Yang, Y.-L. Li, Y.~Zhong, and Y.~Li, \emph{{Thick Brane Split Caused by
			Spacetime Torsion}},
	{\emph{Phys. Rev. D}
		{\bfseries 85} (2012) 084033},
	[{{\ttfamily arXiv:1202.0129}}].
	
	\bibitem{Zhong:2016glr}
	Y.~Zhong, C.-E~Fu, and Y.-X~Liu, \emph{{Cosmological twinlike models with multi scalar fields}},
	{\emph{Sci. China Phys. Mech. Astron.}
		{\bfseries 61} (2018) 90411},
	[{{\ttfamily arXiv:1604.06857}}].
	
	\bibitem{Menezes:2014bta}
	R.~Menezes, \emph{{First Order Formalism for Thick Branes in Modified
			Teleparallel Gravity}},
	{\emph{Phys. Rev. D}
		{\bfseries 89} (2014) 125007},
	[{{\ttfamily arXiv:1403.5587}}].
	
	
	\bibitem{Guo:2015qbt}
	W.-D. Guo, Q.-M. Fu, Y.-P. Zhang, and Y.-X. Liu, \emph{{Tensor perturbations of
			$f(T)$-branes}},
	{\emph{Phys. Rev. D}
		{\bfseries 93} (2016) 044002},
	[{{\ttfamily
			arXiv:1511.07143}}].
	
	\bibitem{Wang:2018jsw}
	J.~Wang, W.-D. Guo, Z.-C. Lin, and Y.-X. Liu, \emph{{Braneworld in $f(T)$
			Gravity Theory with Noncanonical Scalar Matter Field}},
	{\emph{Phys. Rev. D}
		{\bfseries 98} (2018) 084046},
	[{{\ttfamily
			arXiv:1808.00771}}].
	
	\bibitem{DavoodSadatian:2018fss}
	S.~Davood~Sadatian and S.~M. Hosseini, \emph{{Generalization of the
			Randall-Sundrum Model Using Gravitational Model $F(T, \Theta)$}},
	{\emph{Adv. High Energy Phys.}
		{\bfseries 2018} (2018) 2164764},
	[{{\ttfamily
			arXiv:1811.09663}}].
	
	\bibitem{Bamba:2013fta}
	K.~Bamba, S.~Nojiri, and S.~D. Odintsov, \emph{{Effective $F(T)$ gravity from
			the higher-dimensional Kaluza-Klein and Randall-Sundrum theories}},
	{\emph{Phys. Lett. B}
		{\bfseries 725} (2013) 368},
	[{{\ttfamily arXiv:1304.6191}}].
	
	\bibitem{Atazadeh:2014joa}
	K.~Atazadeh and A.~Eghbali, \emph{{Brane cosmology in teleparallel and $f(T)$
			gravity}}, {\emph{Phys.
			Scripta} {\bfseries 90} (2015) 045001},
	[{{\ttfamily arXiv:1406.0624}}].
	
	\bibitem{Correa:2015qma}
	R.~A.~C. Correa and P.~H. R.~S. Moraes, \emph{{Configurational entropy in
			$f\,(R,T\,)$ brane models}},
	{\emph{Eur. Phys. J. C}
		{\bfseries 76} (2016) 100},
	[{{\ttfamily
			arXiv:1509.00732}}].
	
	\bibitem{Guo:2018tpo}
	W.-D. Guo, Y.~Zhong, K.~Yang, T.-T. Sui, and Y.-X. Liu, \emph{{Thick brane in
			mimetic $f(T)$ gravity}},
	{\emph{Phys. Lett. B}
		{\bfseries 800} (2020) 135099},
	[{{\ttfamily
			arXiv:1805.05650}}].
	
	\bibitem{Liu:2008pi}
	Y.-X. Liu, L.-D. Zhang, L.-J. Zhang, and Y.-S. Duan, \emph{{Fermions on Thick
			Branes in Background of Sine-Gordon Kinks}},
	{\emph{Phys. Rev. D}
		{\bfseries 78} (2008) 065025},
	[{{\ttfamily arXiv:0804.4553}}].
	
	\bibitem{Kiritsis:2001bc}
	E.~Kiritsis, N.~Tetradis, and T.~N.~Tomaras,
	\emph{{Induced brane gravity: Realizations and limitations}},
	{\emph{JHEP}  {\bfseries 08} (2001) 012},
	[{{\ttfamily arXiv:hep-th/0106050}}].
	
	
	\bibitem{Tan:2016vwu}
	W.-H.~Tan, S.-Q.~Yang, C.-G.~Shao, J.~Li, A.-B.~Du, B.-F.~Zhan, Q.-L.~Wang, P.-S.~Luo, L.-C.~Tu, and J.~Luo,
	\emph{{New Test of the Gravitational Inverse-Square Law at the Submillimeter Range with Dual Modulation and Compensation}},
	{\emph{Phys. Rev. Lett.}  {\bfseries116} (2016) 131101}.
	
	\bibitem{Lee:2020zjt}
	J.~G.~Lee, E.~G.~Adelberger, T.~S.~Cook, S.~M.~Fleischer, and B.~R.~Heckel,
	\emph{{New Test of the Gravitational $1/r^2$ Law at Separations down to 52 $\mu$m}},
	{\emph{Phys.\ Rev.\ Lett.}
		{\bfseries 124} (2020) 101101},
	[{{\ttfamily arXiv:2002.11761}}].
	
	\bibitem{Tan:2020vpf}
	W.-H.~Tan, A.-B.~Du, W.-C.~Dong, S.-Q.~Yang, C.-G.~Shao, S.-G.~Guan, Q.-L.~Wang, B.-F.~Zhan, P.-S.~Luo, L.-C.~Tu, and J.~Luo,
	\emph{{Improvement for Testing the Gravitational Inverse-Square Law at the Submillimeter Range}},
	{\emph{Phys. Rev. Lett.} {\bfseries 124} (2020) 051301}.
	
	
	
\end{thebibliography}
\end{document}